\documentclass[12pt]{article}
\usepackage{amssymb}
\usepackage{graphicx}
\usepackage{amsmath}
\usepackage{array}

\def\thline{\noalign{\hrule height 1pt}}
\def\tvline{\vrule width 1pt} 

\newcommand {\beq}{\begin{eqnarray}}
\newcommand {\eeq}{\end{eqnarray}}
\def\p{\partial}
\newcommand{\NF}{N_{\rm F}}
\newcommand{\NC}{N_{\rm C}}

\newcommand{\hs}[1]{\hspace{#1 mm}}
\newcommand{\bpm}{\begin{pmatrix}}
\newcommand{\epm}{\end{pmatrix}}

\newcommand{\Sym}{\mathfrak{S}}

\begin{document}
\thispagestyle{empty}
\begin{flushright}
TIT/HEP--548 \\
RIKEN-TH-63 \\
{\tt hep-th/0601181} \\
January, 2006 \\
\end{flushright}
\vspace{3mm}

\begin{center}
{\LARGE \bf 
Non-Abelian Vortices on Cylinder
} \\ 
\vspace{0.5cm}
{\Large
---  \  Duality between vortices and walls  \ ---
}
\\[12mm]
\vspace{5mm}

{\normalsize\bfseries
Minoru~Eto$^1$, 
Toshiaki~Fujimori$^1$, 
Youichi~Isozumi$^1$,
Muneto~Nitta$^1$, \\
 Keisuke~Ohashi$ ^1$, 
 Kazutoshi~Ohta $^2$
and
 Norisuke~Sakai $^1$}
\footnotetext{
e-mail~addresses: \tt
meto,
fujimori,
isozumi,
nitta,
keisuke@th.phys.titech.ac.jp;
k-ohta@riken.jp;\\
nsakai@th.phys.titech.ac.jp
}

\vskip 1.5em

{ \it $^1$Department of Physics, Tokyo Institute of 
Technology \\
Tokyo 152-8551, JAPAN  \\[2mm]
$^2$
Theoretical Physics
Laboratory\\
 The Institute of Physical and Chemical Research
 (RIKEN) \\
Wako 351-0198, JAPAN
 }
\vspace{12mm}

\abstract{

We investigate vortices on a cylinder in supersymmetric 
non-Abelian 
gauge theory with 
hypermultiplets in the 
fundamental representation.
We identify moduli space of periodic vortices and find 
that a pair of wall-like objects appears as the vortex 
moduli is varied. 
Usual domain walls also can be obtained from the single 
vortex on the cylinder by introducing a twisted boundary 
condition. 
We can understand these phenomena as a T-duality among 
D-brane configurations in type II superstring theories. 
Using this T-duality picture, we find a 
one-to-one correspondence between the moduli space of 
non-Abelian vortices and that of kinky D-brane 
configurations for domain walls. 

}

\end{center}

\vfill
\newpage
\setcounter{page}{1}
\setcounter{footnote}{0}
\renewcommand{\thefootnote}{\arabic{footnote}}

\section{Introduction}\label{intro}

Solitons play extremely important role in both gauge theory 
and string theory to understand their non-perturbative 
dynamics.
Especially, Bogomol'nyi-Prasad-Sommerfield (BPS) 
solitons \cite{Bogomolny:1975de} 
in supersymmetric (SUSY) theory, which
preserve a fraction of supercharges \cite{Witten:1978mh}, 
enjoy analytically good properties
due to their integrability. 
It is a very important task to construct an explicit 
solution and moduli space of the solitons in gauge theory, 
although it is very difficult in general. 
Atiyah, Drinfeld, Hitchin, and Manin (ADHM) 
gave a procedure 
for the construction of instanton (self-dual field strength) 
solutions~\cite{Atiyah:1978ri, Corrigan:1983sv}.
Using this construction, we can understand the structure of 
the moduli space of instantons. 
On the other hand, Nahm constructed solutions of 
't\,Hooft-Polyakov monopoles in the the BPS limit 
and investigated their moduli space using 
very similar techniques to the ADHM 
construction of instantons 
\cite{Nahm:1979yw, Nahm:1980jy, Nahm:1983sv}.
The Nahm's construction embodies a basic idea of the 
relationship between instantons with periodic boundary 
condition, called calorons, and monopoles.
This relationship is a kind of the Fourier transformation 
with respect to the periodic direction and can be 
understood as a part of dualities of instantons 
on $T^4$ \cite{Schenk:1986xe, Braam:1988qk}.
The duality between the ADHM and Nahm constructions 
is now regarded as the T-duality among D-brane 
configurations in string theory. 
Namely, if we realize the system of instantons
on $T^4$ as a D0-D4 bound state, 
for example, the T-duality maps it to 
a D1-D3 bound state 
which represents a profile of the solution of 
the Nahm equation \cite{Diaconescu:1996rk, Lee:1997vp}. 
The ADHM or Nahm equation appears as the BPS 
equation of the effective theory on D-branes. 
So string theory provides us a lucid picture on dualities 
between instantons and monopoles. 

Recently BPS solitons with less codimensions have 
attracted much attention; 
both vortices~\cite{Abrikosov:1956sx}--\cite{Alford:1990mk} 
and domain walls~\cite{Abraham:1992vb}--\cite{Hanany:2005bq} 
are BPS solitons existing 
in the Higgs phase of SUSY gauge theory 
with eight supercharges coupled with hypermultiplets.  
Previously known vortices, called the 
Abrikosov-Nielsen-Olesen (ANO) 
vortices~\cite{Abrikosov:1956sx,Taubes:1979tm,Samols:1991ne}, 
in $U(1)$ gauge theory with one Higgs field 
have been recently extended to 
non-Abelian vortices~\cite{Hanany:2003hp}--\cite{Auzzi:2005gr} 
in $U(\NC)$ gauge theory with 
$\NF (\geq \NC)$ fundamental Higgs fields.\footnote{
Another type of vortices in non-Abelian gauge theory 
were also known \cite{Alford:1990mk}. 
}
Especially in Ref.~\cite{Eto:2005yh} 
the moduli space of non-Abelian vortices has been 
completely determined (except for the metric) 
in the case of $\NC = \NF$. 
Domain walls in $U(1)$ gauge 
theory~\cite{Abraham:1992vb}--\cite{Bolognesi:2005zr} 
have also been extended to non-Abelian 
domain walls~\cite{Shifman:2003uh}--\cite{Hanany:2005bq}, 
and their moduli space has been completely determined 
\cite{Isozumi:2004jc,Isozumi:2004va}.  
See \cite{Tong:2005un} as a review on vortices and domain walls.
Like instantons and monopoles, 
vortices and domain walls also have nice 
interpretation by D-brane systems 
in string theory; 
D-brane configurations for vortices 
are given in Refs.~\cite{Hanany:2003hp,Hanany:2004ea} 
whereas domain walls are realized by kinky D-brane 
configurations \cite{Lambert:1999ix,Eto:2004vy,Eto:2005mx}. 
One interesting observation is that 
solutions and moduli space of vortices (domain walls) 
can, roughly speaking, be regarded as a ``half'' of those 
of instantons \cite{Hanany:2003hp,Hanany:2004ea} 
(monopoles \cite{Hanany:2005bq}).
We can find that the vortex/domain wall moduli space 
is a Lagrangian (mid-dimensional) submanifold of 
instanton/monopole one. 
So we naturally expect that there exist a similar duality 
relation between vortices and domain walls. 
However, in contrast to the outstanding results on the 
instanton/monopole duality, 
the relation between vortices and domain walls has not been 
investigated in depth. 
The idea of duality between vortices and domain walls 
in {\it Abelian} gauge theory has been already 
proposed in \cite{Gauntlett:2000de, Tong:2002hi}.

In this paper, 
we extend this idea of vortex/domain-wall duality to 
the case of {\it non-Abelian} gauge theory 
by using a concrete solution
of vortices on a cylinder ${\bf R}\times S^1$. 
To this end we use the method of the moduli matrix, 
which has been  
introduced by a part of present authors for study of 
domain walls~\cite{Isozumi:2004jc,Isozumi:2004va,Eto:2005wf} 
and has been extended to vortices \cite{Eto:2005yh} and 
other composite solitons like 
wall-wall junction \cite{Eto:2005cp,Eto:2005fm}, 
monopole-vortex-wall junction \cite{Isozumi:2004vg} and  
instanton-vortex junction \cite{Eto:2004rz}.
(See Refs.~\cite{rev-modulimatrix,rev-modulimatrix2} 
for a review of this method.)
In particular, it has been shown in Ref.~\cite{Eto:2005yh} 
that the moduli parameters of non-Abelian vortices on a 
plane ${\bf R}^2$ are encoded in the moduli matrix whose 
elements are polynomials holomorphic with respect to the 
complex variable made of codimensions of vortices. 
On the other hand, the vortex solution on ${\bf R}\times S^1$ 
possesses a periodic boundary condition due to the 
compactified direction along $S^1$ with radius $R$. 
As a result the moduli matrix giving a vortex solution 
is written holomorphically in terms of trigonometric
functions, which reflects the periodicity, instead of 
polynomials of the holomorphic coordinate. 
We will see a phenomenon that the periodic vortices in 
the fundamental region join with each other and form 
two wall-like objects, where the tension (energy density) 
is of the order of Kaluza-Klein (KK) mass, namely $1/R$,
if a typical size of vortices exceeds the radius of $S^1$.
If we view this phenomenon from a dual cylinder side 
${\bf R}\times \hat{S}^1$, where $\hat{S}^1$ has a dual 
radius $\hat{R}=1/R$, the profile of the domain wall 
appears as a kink of the Wilson lines. 
We also produce domain walls with tension much smaller 
than $1/R$ by turning on a twisted boundary condition 
to Higgs fields along $S^1$, which is referred to as 
Scherk-Schwarz (SS) compactification in our context. 
Using these relations of solutions, 
we can explicitly determine the correspondence between 
moduli parameters of vortices and domain walls.

We also interpret the above vortex/wall duality  
from the string theoretical point of view as a kind of 
T-duality among D-brane configurations for 
vortices~\cite{Hanany:2003hp,Hanany:2004ea} 
and domain walls~\cite{Eto:2004vy}.
The profile of the Wilson line in the vortex 
configuration represents directly 
a shape of kinky D-brane on the dual cylinder.
It is easy to read meanings of the moduli parameters 
of vortices from the kinky D-brane
configuration if we use the D-brane picture. 
The D-branes help us to investigate the properties of 
vortices and support 
our observation on the vortex/domain wall duality.

The organization of the paper is as follows:
In Sec.~\ref{sec:vw}, 
we give a brief review of vortices and domain walls. 
In particular in the first subsection 
we discuss in detail a characteristic property of 
semi-local vortices\footnote{
Semi-local vortices \cite{Vachaspati:1991dz} exist 
in $U(\NC)$ gauge theory with $\NF (> \NC)$ fundamental 
Higgs fields. 
They reduce sigma-model lumps \cite{Ward:1985ij}  
in the strong gauge coupling limit. 
}; 
When they coincide a hole appears.
This will become a key observation for duality between 
vortices and walls. 
We explain that the $d=1+1$ gauge theory with massive 
hypermultiplets admitting domain walls 
can be obtained by the Scherk-Schwarz dimensional reduction 
from the $d=2+1$ gauge theory with massless hypermultiplets 
admitting vortices.
This is the first evidence of duality between vortices and 
domain walls. 
In Sec.~\ref{sec:v2w}, we construct vortices on a cylinder 
${\bf R}\times S^1$ and show that there exists a 
transition between vortices and wall-like objects 
by changing the moduli parameters. 
Then, we impose a twisted boundary condition along the 
compactified direction $S^1$. 
We calculate the Wilson loops around $S^1$ in a calculable 
example and see that the usual domain walls appear as kinks 
of the Wilson line on the dual circle $\hat{S}^1$. 
This gives nothing but the position of the D-brane in 
compactified direction.
In Sec.~\ref{sec:brane}, we show that the brane 
configuration for vortices 
and the one of domain walls are related by T-duality. 
We begin with the brane configuration for vortices given 
in \cite{Hanany:2003hp}. 
The brane configuration for domain walls is obtained as 
the T-dual picture of this configuration.
In Sec.~\ref{MSVW} 
we investigate the structure of vortex moduli space 
in terms of domain wall moduli space by applying this duality. 
In Sec.~\ref{Disc}, we give conclusion and discussion.
Possibility of new kind of field theory supertubes and 
similarity with tachyon condensation are pointed out.

\section{Vortices and Walls}\label{sec:vw}
\subsection{Vortices}\label{sub:vortex}

First, we briefly review the construction of vortices 
in $(2+1)$-dimensional $U(\NC)$ gauge theory with eight 
supercharges \cite{Eto:2005yh}. 
We consider a model with $\NF (\geq \NC)$ massless 
hypermultiplets in the fundamental representation. 
The physical fields in the vector multiplet are a 
$U(\NC)$ gauge field $W_M\ (M = 0,1,2)$, three real 
adjoint scalars $\Sigma^I\ (I = 1,2,3)$ and 
their fermionic partners. 
The hypermultiplets consist of $SU(2)_R$ doublet of 
complex scalars 
$H^{irA}\ (i=1,2, \hs{2} r=1,2,\cdots,\NC, \hs{2} A=1,2,\cdots,\NF)$ 
and their fermionic partners. 
We express $\NC \times \NF$ matrices of the hypermultiplets 
by $H^i$. The bosonic Lagrangian takes the form of
\beq
\mathcal L_{2+1} = {\rm Tr} \bigg[- \frac{1}{2g^2} F_{MN}F^{MN} 
+ \frac{1}{g^2}\mathcal D_M \Sigma^I \mathcal D^M \Sigma^I 
+\mathcal D_M H^i (\mathcal D^M H^i)^\dagger \hs{10}\nonumber \\ 
- \frac{1}{g^2}[\Sigma^I,\Sigma^{J}]^2 
- H^i (H^i)^\dagger \Sigma^I \Sigma^I 
- \frac{1}{g^2}(Y^a)^2\bigg],
\label{eq:vaction}
\eeq
where the trace is taken over the color indices and 
$Y^a = \frac{g^2}{2}(c^a {\mathbf 1}_{\NC} 
- (\sigma^a)^j{}_i H^i H^{j\dagger}) 
\ (a = 1,2,3)$. 
Here the real parameters $c^a$ are called the 
Fayet-Iliopoulos (FI) parameters 
which we fix $c^a = (0,0,c>0)$ by using the $SU(2)_R$ 
rotation and $g$ is the gauge coupling\footnote{
For simplicity we choose identical gauge couplings for the 
$U(1)$ and $SU(N_{\rm C})$ gauge groups. 
}. 
The covariant derivatives and the field strength are 
defined by 
$\mathcal D_M \Sigma^I= \p_M\Sigma^I + i[W_M,\Sigma^I], 
\hs{2} \mathcal D_M H^i = \p_MH^i + iW_MH^i$ and 
$F_{MN} = -i[\mathcal D_M, 
\mathcal D_N] = \p_M W_N - \p_N W_M + i[W_M,W_N]$. 
Taking $\NC = \NF = 1$ this Lagrangian reduces to $U(1)$ 
gauge theory admitting Abelian vortices, 
called the Abrikosov-Nielsen-Olesen (ANO) vortices 
\cite{Abrikosov:1956sx}. 

The conditions for supersymmetric vacua are given by
\beq
H^1 (H^1)^\dagger - H^2(H^2)^\dagger 
= c {\mathbf 1}_{\NC}, 
\hs{2} H^1(H^2)^\dagger =0, 
\hs{2} \Sigma^I H^i =0, \hs{2} [\Sigma^I,\Sigma^{J}] = 0.
\label{eq:vacuum1}
\eeq
Since the first equation and the third equation 
require $\Sigma^I$ to vanish for all $I$, the vacua are 
in the Higgs branch with completely 
broken $U(\NC)$ gauge symmetry. 
For $\NC = \NF$, the vacuum conditions 
(\ref{eq:vacuum1}) determine the unique vacuum 
$H^1= \sqrt{c}{\mathbf 1}_{\NC}, \hs{2} H^2=0$, 
called the color-flavor locking phase. 
For $\NF > \NC$, the moduli space of vacua is the 
cotangent bundle over the complex Grassmannian, 
$T^{\ast} G_{\NF,\NC} 
= T^{\ast}[SU(\NF)/(SU(\NF-\NC) \times SU(\NC) \times U(1))]$ 
(see, e.g., \cite{Arai:2003tc}). 

Let us consider BPS solutions which have non-trivial 
configurations of $F_{12}$ and $H^1$
while all the other fields vanish. 
We assume that the solutions are static and depend on
$x^1$ and $x^2$ only. 
For later convenience, we shall define a complex 
coordinate $z=x^1+ix^2$ and the associated covariant 
derivative as 
$\mathcal D_z = (\mathcal D_1 + i \mathcal D_2)/2$.
The energy density is bounded from below by the 
Bogomol'nyi completion as
\beq
\mathcal E &=&{\rm Tr} \bigg[ \frac{1}{g^2} F_{12} F_{12} 
+ \mathcal D_{m} H(\mathcal D_m H)^{\dagger} 
+ \frac{g^2}{4}(c{\mathbf 1}_{\NC}-H H^{\dagger})^2  \bigg] 
\nonumber \\ 
&=& {\rm Tr} \bigg[ \frac{1}{g^2}\left(F_{12} 
+ \frac{g^2}{2}(c{\mathbf 1}_{\NC}-H H^{\dagger})\right)^2 
+ 4 \bar{\mathcal D}_{ z} H (\bar{\mathcal D}_{ z} H)^{\dagger} 
- c F_{12} \bigg] \nonumber \\
&\geq& -c {\rm Tr} F_{12}
\label{eq: energy}
\eeq
Here we simply denote the hypermultiplet scalars $H^1$ as $H$. 
The BPS solutions saturate the BPS energy bound and 
satisfy the following BPS equations
\beq
F_{12} = - \frac{g^2}{2} (c {\mathbf 1}_{\NC} -H H^{\dagger} ) ,
\quad
\bar{\mathcal D}_{ z} H = 0.
\label{eq:BPS1}
\eeq
Note that these equations can also be derived
by requiring half of the supersymmetries to be preserved. 
The second equation of the Eq.(\ref{eq:BPS1}) can be formally 
solved as \cite{Eto:2005yh}
\beq
 \bar W_{z} = -i S^{-1} \overline \p_z S, 
\hs{5} H = S^{-1}(z,\bar z) H_0(z),
 \label{sol-vol}
\eeq
where $S(z, \bar z)$ is an $\NC \times \NC$ non-singular 
matrix function and $H_0(z)$ is an $\NC \times \NF$ 
matrix whose components are arbitrary 
holomorphic functions with respect to $z$.
We call the matrix $H_0$ moduli matrix for vortices 
because constants in $H_0$ are moduli parameters of this 
system.
Once a $H_0$ is given, the matrix $S$ is determined 
by the first equation of Eq.(\ref{eq:BPS1}) up to 
$U(\NC)$ gauge transformations. 
To this end, it is useful to define a gauge invariant 
$\NC \times \NC$ matrix
\beq 
\Omega (z,\bar{z})= S(z,\bar{z}) S^{\dagger}(z,\bar{z}).
\eeq 
Then the first of Eq.(\ref{eq:BPS1}) can be rewritten as 
\cite{Eto:2005yh}
\beq
\overline \p_z (\p_z \Omega \Omega^{-1}) 
= \frac{g^2c}{4}({\mathbf 1}_{\NC} - \Omega_0 \Omega^{-1}), 
\hs{5} 
\Omega_0 \equiv \frac{1}{c} H_0 H_0^{\dagger}, 
\label{eq:master}
\eeq
which we call the master equation for vortices.
For finite energy solutions, the configurations must 
approach a point of the moduli space of vacua  at spatial 
infinity, that is 
$H H^{\dagger} \rightarrow c {\mathbf 1}_{\NC}$ 
as $ |z| \rightarrow \infty$. 
In terms of $\Omega$, this boundary condition can be 
rewritten as $\Omega \rightarrow \Omega_0$ at spatial 
infinity. 
For given $H_0$, it is thought that the equation 
(\ref{eq:master}) uniquely determine 
$\Omega$ without any additional constants of integration. 
Therefore, we expect that all moduli parameters are 
contained in the moduli matrix $H_0$. 
There exists an equivalence relation which we call 
$V$-equivalence relation
\beq
(H_0(z), \ S(z, \bar z)) \ \sim \ V(z)(H_0(z),\ S(z, \bar z)), 
\label{V-equiv-vol}
\eeq
where $V(z) \in GL(\NC, {\mathbf C})$ is a non-singular matrix 
function whose components are holomorphic with respect to $z$. 
It is obvious  from Eq.(\ref{eq:BPS1}) 
that moduli matrices belonging to the same equivalence class 
give the same physical quantities. 
Then the moduli space for $k$-vortices 
$\mathcal M^k_{\rm v}$ can be identified with a quotient 
space defined by the equivalence relation $\sim$, 
\beq 
&\mathcal M^k_{\rm v} \cong \mathcal G / \sim, 
& \label{tm_vortex}\label{ms:vortex}\\ 
&\mathcal G \equiv \{H_0| H_0: 
\mathbf C \rightarrow M(\NC \times \NF,
\ \mathbf C),\ \overline \p_z H_0 = 0,
\ \det\Omega_0=\mathcal O(|z|^{2k})\},
& \nonumber \\ & H_0(z) \sim V(z) H_0(z),
 \hs{5} V(z) \in GL(\NC, \mathbf C), 
\hs{5} \overline \p_z V(z) = 0.& \nonumber
\eeq

The energies of the BPS configurations are characterized by the vorticity
$k  \in {\mathbf Z}$ as
\beq
T_{\rm v} = -c \int dx^1 dx^2\ {\rm Tr} F_{12} 
= c \oint_\infty dz \ \p_z {\rm log }\hs{1}{\rm det} 
\Omega_0 = 2\pi c k,
\label{top:vortex}
\eeq 
where we used 
${\rm Tr}  F_{12} =  - \overline \p_z \p_z \log \det \Omega.$
Then the highest power of $|z|^2$ in ${\rm det} \Omega_0$
determines the vorticity $k$. 

As a simple example, let us consider the ANO vortices 
in the $\NC=\NF=1$ model.
In this case, the moduli matrix is just a holomorphic 
polynomial
\beq
H_0 = \sqrt c \prod_{i=1}^{k} (z-p_i)
\label{mm_ano}
\eeq
giving the $k$-vortex configurations. 
The complex constant parameter $p_i$ represents the position 
of the $i$-th vortex.
The master equation (\ref{eq:master}) with this moduli 
matrix is equivalent under a redefinition 
$c |\Omega| = |H_0|^2 e^{-\xi(z,\bar z)}$ 
to the so-called Taubes equation 
for the ANO vortices, existence and uniqueness of whose 
solutions were proved \cite{Taubes:1979tm}. 
There is a fundamental quantity $g\sqrt c$ which is only 
one parameter with mass dimension in this system, see 
Eq.(\ref{eq:master}). 
The size of the ANO vortices is of order $1/(g\sqrt c)$.

Let us next give another example of the semi-local vortex
in the case of $\NF=2,\ \NC=1$.
The $k$-vortex solutions are generated by the moduli matrix
\beq
H_0 = \sqrt c \left(a \prod_{i=1}^{k-1}(z-q_i),
\  \prod_{i=1}^{k} (z-p_i)\right),
\label{mm_semi}
\eeq
where we have fixed the boundary condition\footnote{
One should note that $H \rightarrow \sqrt{c} \bpm 0,&1 \epm$ at 
$|z| \rightarrow \infty$ as illustrated in 
Eq.~(\ref{eq:hole_coincident_vortices}) 
for the strong coupling limit, whereas the moduli matrix does 
not reduce to $H_0 \rightarrow \sqrt{c} \bpm 0,&1 \epm$ at 
$|z| \rightarrow \infty$ even if the $V$-equivalence relation 
(\ref{V-equiv-vol}) is used. 
}
as 
$H \rightarrow \sqrt{c} \bpm 0,&1 \epm$ at spatial 
infinity $|z| \rightarrow \infty$.
The moduli parameters $\{p_1,p_2,\cdots,p_k\}$ 
have one to one correspondence with the positions of 
the $k$-vortices and the parameters $a$ and 
$\{q_1,q_2,\cdots,q_{k-1}\}$ 
parameterize the total and relative sizes of 
$k$-vortices, respectively.
Comparing the moduli matrix (\ref{mm_semi}) for the 
semi-local vortices with that (\ref{mm_ano}) for the 
ANO vortices, we see that the semi-local vortices 
have size moduli parameters $(a,q_1,\cdots,q_{k-1})$ 
in addition to the position moduli. 
These size parameters will play an important role when 
we discuss the duality between the vortices
and the domain walls in the following sections.

\begin{figure}[ht]
\begin{center}
\includegraphics[width=15cm]{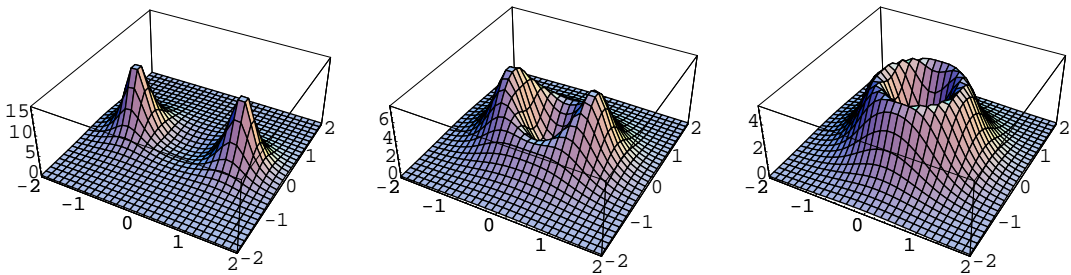}
\caption{
Two separated lumps (left) approach each other (middle). 
A ring appears at the coincident limit (right). 
}
\label{hole}
\end{center}
\end{figure}

One of the interesting properties of the semi-local 
vortices is that a vacuum different from 
$H=\sqrt c(0,\ 1)$ at infinity appears inside a vortex 
exhibiting a hole (ring) when two or more vortices coincide.
This property will be very important for 
the duality between vortices and domain walls.
In order to understand this, we shall 
first take the strong
gauge coupling limit\footnote{
In this limit the gauged linear sigma model 
(\ref{eq:vaction}) reduces to the hyper-K\"ahler 
non-linear sigma model whose target space is 
the Higgs branch $T^* G_{\NF,\NC}$~\cite{Arai:2003tc}.
At the same time the semi-local vortices 
reduce to the Grassmann 
sigma-model lumps~\cite{Ward:1985ij}.
}
$g^2\to\infty$ in which the master equation 
(\ref{eq:master}) reduces to an algebraic equation 
and can be solved concretely.
Then we can exactly solve
the BPS equation (\ref{eq:BPS1}).
Let us consider the two lumps given by
$H_0=\sqrt c (a^2,\ (z-p_1)(z-p_2))$. 
This gives the following solution
\beq
H = \sqrt c \left( \frac{a^2}{\rho(z)},\ 
\frac{(z-p_1)(z-p_2)}{\rho(z)}\right),\qquad
{\cal E} = \frac{c|a|^4 |2z-(p_1+p_2)|^2}{\rho(z)^4},
\label{eq:hole_coincident_vortices}
\eeq
with $\rho(z) \equiv \sqrt{|a|^4+|z-p_1|^2|z-p_2|^2}$.
The configuration approaches the vacuum 
$H = \sqrt c \left(0,\ 1\right)$ at the spatial infinity.
Notice that the energy density vanishes 
at the center of mass $z = (p_1+p_2)/2$.
Especially, it becomes the vacuum 
$H=\sqrt c\left(1,\ 0\right)$ 
when the two vortices coincide. 
Then the energy density exhibits 
a ring around the coincident position of these lumps 
because the axial symmetry is recovered there.
Important point is 
that the vacuum appearing inside the coincident position of 
the lumps (inside the hole) 
is different from the outside one.
We show the profiles of the energy densities for two lumps 
in Fig.~\ref{hole}.

\begin{figure}[ht]
\begin{center}
\includegraphics[height=5cm]{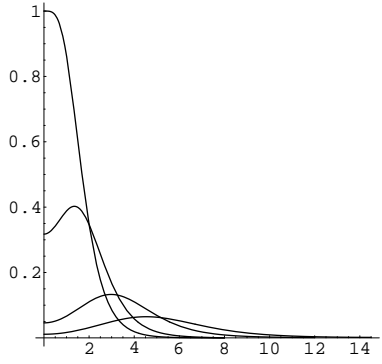}
\caption{
Energy density of two coincident semi-local vortices 
in the case of $\NC=1$ and $\NF=2$. 
The horizontal axis denotes the radius from the coincident 
point and the vertical axis denotes the energy density. 
Energy densities 
with various values of the size modulus $|a|$ 
are plotted. 
It has a huge central peak for $a=0$, whereas 
a central hole with a ring structure develops 
for larger $|a|$. 
}
\label{semi_hole}
\end{center}
\end{figure}

A ring structure appears even in finite gauge coupling 
$(g^2<\infty)$ 
when two semi-local vortices coincide. 
However the energy density does not exactly vanish 
and no vacuum appears
at the center of the ring 
unlike the case of the lumps.
We show the energy density of 
two coincident semi-local vortices 
with various size modulus $|a|$ 
in the Fig.~\ref{semi_hole}.
The energy density at the center of the ring increases 
as the total size $|a|$ of the 
semi-local vortices becomes small.
This is understood as follows.
In the theory with the finite gauge coupling there is the 
fundamental size $1/(g\sqrt c)$ 
which is the size of the ANO vortex as we mentioned above.
When the size parameter of the semi-local vortex
is much smaller than the ANO size ($|a| \ll 1/(g\sqrt c)$),
the semi-local vortex is almost an ANO vortex. 
Actually, the configuration with $|a|=0$ is identical with 
the ANO vortex.
On the other hand, the semi-local vortex becomes 
a lump-like solution with a peak around $|z| \sim |a|$ 
when $|a| \gg 1/(g\sqrt c)$. 
In the limit where the size $|a|$ is taken to infinity, 
the configuration reduces to the vacuum $H = \sqrt c(1,\ 0)$.

\subsection{Domain Walls}\label{sub:wall}

Let us next turn our attention to the construction of 
domain walls by using the moduli 
matrix~\cite{Isozumi:2004jc,Isozumi:2004va}.
Similarly to the BPS vortex in 2+1 dimensions in the 
previous subsection, 
we again consider the supersymmetric $U(\NC)$ gauge theory 
with eight supercharges. 
We consider this theory in spacetime dimension 1+1 and
turn on the non-degenerate real masses 
\beq
M = {\rm diag}\left(m_1,m_2,\cdots,m_{\NF}\right),\quad
\left(m_A > m_{A+1}\right)
\eeq
for the 
$\NF (>\NC)$ hypermultiplets. 
Due to these
hypermultiplet masses,
the flavor symmetry $SU(\NF)$ is explicitly broken to its 
maximal Abelian subgroup 
$U(1)^{\NF-1}$.  
The bosonic fields in the vector multiplet are 
the gauge field $W_m,(m = 0,1)$ and four real adjoint scalars 
$\Sigma^I$ $(I=1,2,3)$ and $\hat{\Sigma}$. 
We denote $\hat H^i$ as the hypermultiplet scalars
and $(\hat c,\hat g)$ as the FI parameter and the gauge coupling
in 1+1 dimensions, respectively.
Then, the bosonic part of Lagrangian takes the form of
\beq
\mathcal L_{1+1} 
=  \hs{1}  {\rm Tr} \bigg[ \hs{-3} &-& \hs{-3} \frac{1}{2 \hat g^2} F_{mn}F^{mn}+ \frac{1}{\hat g^2}\mathcal D_m \Sigma^I \mathcal D^m \Sigma^I 
+ \frac{1}{\hat g^2}\mathcal D_m \hat \Sigma \mathcal D^m \hat \Sigma 
\\ \nonumber 
&+& \hs{-2} \mathcal D_m \hat H^i (\mathcal D^m \hat H^i)^\dagger
- \frac{1}{2 \hat g^2}[\Sigma^I,\Sigma^{J}]^2 
- \frac{1}{\hat g^2}[\Sigma^I,\hat{\Sigma}]^2 
- \hat H^i (\hat H^i)^\dagger \Sigma^I \Sigma^I 
\nonumber \\
&+& \hs{-2} (\hat{\Sigma} \hat H^i-\hat H^i M)
(\hat{\Sigma} \hat H^i - \hat H^i M)^{\dagger} 
- \frac{\hat g^2}{4}(\hat c^a {\mathbf 1}_{\NC} - 
(\sigma^a)^j{}_i \hat H^i \hat H^{j\dagger})^2\bigg]. 
\label{eq:waction}
\eeq
The most points in the vacuum manifold $T^* G_{\NF,\NC}$
of the massless theory 
are lifted by the nondegenerate hypermultiplet masses 
except for the $_{\NF}C_{\NC}$ discrete points, given by  
\beq
\hat H^{1rA} = \sqrt{\hat c} \delta^A{}_{A_r}, 
\hs{5} \hat H^2=\Sigma^I=0, 
\hs{5} \hat{\Sigma} 
= {\rm diag}(m_{A_1},m_{A_2},\cdots,m_{A_{\NC}}),
\label{eq:discrete_vacua}
\eeq
where the first equation means that only one flavor 
component of each (color) row must be non-vanishing. 
Therefore the vacua are discrete and labeled by sets of the 
flavors $\langle A_1,A_2,\cdots,A_{\NC}\rangle$. 
These are also called color-flavor locking vacua. 

The BPS equations for the domain walls interpolating 
between these discrete vacua are of the form 
\beq
\mathcal D_1 \hat\Sigma 
= \frac{\hat g^2}{2}(\hat c {\mathbf 1}_{\NC} 
- \hat H \hat H^{\dagger}),\quad
\mathcal D_1 \hat H = -\hat\Sigma \hat H + \hat H M.
\label{eq:BPS2}
\eeq
We solve these first order equations by imposing the 
boundary condition 
$\langle A_1,A_2,\cdots,A_{\NC}\rangle$ at $x^1=+\infty$
and 
$\langle B_1,B_2,\cdots,B_{\NC}\rangle$ at $x^1=-\infty$.
Here we set $A_r \le B_r$, so that $m_{A_r} \ge m_{B_r}$.
The energy of the domain wall is given by the topological 
charge as 
\beq
T_{\rm w} = \int_{-\infty}^{\infty} dx^1
\  \hat c \ {\rm Tr} (\mathcal D_1 \hat\Sigma) 
= \hat c[{\rm Tr} \hat\Sigma ]^{\infty}_{-\infty} 
= \hat c \left( \sum_{k=1}^{\NC} m_{A_k} 
- \sum_{k=1}^{\NC} m_{B_k} \right)
> 0.
\label{top:wall}
\eeq
As in the case of the vortices, 
the second equation of (\ref{eq:BPS2}) can be solved
by introducing the non-singular 
$\NC\times\NC$ matrix $\hat S$ 
and the $\NC\times\NF$ constant matrix $\hat H_0$ as 
\cite{Isozumi:2004jc,Isozumi:2004va}
\beq
\hat\Sigma + iW_1 =\hat S^{-1} \p_1 \hat S, 
\ \ \ \ 
\hat H(x^1) = \hat S^{-1}(x^1) \hat H_0 e^{Mx^1}. 
\label{sol-wall}
\eeq
We call $\hat H_0$ the moduli matrix for the domain walls. 
Notice that $\hat S(x^1)$ depends on $x^1$ only while
the moduli matrix $\hat H_0$ is a constant matrix.
Defining an $\NC \times \NC$ gauge invariant matrix
\beq 
\hat\Omega(x^1) = \hat S(x^1) \hat S^{\dagger}(x^1),
\eeq
the first equation of Eq. (\ref{eq:BPS2}) can be 
rewritten as \cite{Isozumi:2004jc,Isozumi:2004va}
\beq
 \p_1(\p_1 \hat\Omega \hat\Omega^{-1}) 
 = \hat g^2 \hat c({\mathbf 1}_{\NC} 
- \hat\Omega_0 \hat\Omega^{-1}),
 \ \ \ 
 \hat\Omega_0\equiv \frac{1}{\hat c} 
\hat H_0 e^{2M x^1} \hat H^{\dagger}_0 ,
  \label{master-eq-wall}
\eeq
which we call the master equation for domain walls. 
Once the solution $\hat\Omega(x^1)$ of this equation is 
obtained, we can calculate all the physical quantities. 
In the strong gauge coupling limit $g^2 \to \infty$ this 
equation reduces to an algebraic equation, which can be 
solved immediately and uniquely. 
For finite gauge coupling, uniqueness and existence of 
the solution of the master equation (\ref{master-eq-wall}) 
were proved for $\NC=1$ \cite{Sakai:2005kz} 
and are consistent with the index theorem \cite{Sakai:2005sp} 
for arbitrary $\NC (< \NF)$.

The moduli space of the domain walls is obtained by the 
same way with the vortices. 
First note that there is the $\hat V$-equivalence relation
\beq
(\hat H_0, \ \hat S(x^1)) \ 
\sim \ \hat V (\hat H_0, \ \hat S(x^1))
 \label{V-equiv-wall}
\eeq
where  $\hat V \in GL(\NC,{\mathbf C})$ is a mere 
constant matrix. 
Therefore the total moduli space of BPS domain walls 
is the complex Grassmannian:
\beq
 \mathcal M_{\rm w} 
\cong \{\hat H_0| \hat H_0 \sim \hat V \hat H_0, 
 \hat V \in GL(\NC,\mathbf C)\} \cong G_{\NF,\NC} = 
 {SU(\NF) \over SU(\NF-\NC) \times SU(\NC) \times U(1)}
.
 \label{ms:wall}
\eeq
Since we did not enforce any boundary condition, 
unlike the one $\det \Omega_0 = {\cal O}(|z|^{2k})$ for 
vortices, we obtain the total moduli space including 
all topological sectors connected together properly.
\footnote{
There exists a homeomorphism between the total moduli 
space $G_{\NF,\NC}$ 
and (a base manifold of) the moduli space 
$T^* G_{\NF,\NC}$ of the Higgs branch of vacua 
of the corresponding massless theory.
This is not a coincidence: 
it was shown in Ref.~\cite{Eto:2005wf} 
that the total moduli space of domain walls is always 
the union of special Lagrangian submanifolds 
of the Higgs branch moduli space of the corresponding 
massless theory. 
} 
We can obtain each topological sector 
by enforcing a proper boundary condition. 

\subsection{Scherk-Schwarz dimensional reduction}
\label{sub:ss}
The massive action in 1+1 dimensions in 
Sec.~\ref{sub:wall} can be obtained 
from the massless action in 2+1 dimensions in 
Sec.~\ref{sub:vortex}
by the the Scherk-Schwarz (SS) dimensional 
reduction.\footnote{
Since we twist the flavor symmetry commuting with all SUSY, 
SUSY is preserved, 
unlike the SS reduction with twisting $SU(2)_R$.
}
In the SS dimensional reduction, the $x^2$ direction is 
compactified on $S^1$ with radius 
$2 \pi R\ (\ll 2 \pi/m_A)$ and we impose a twisted 
boundary condition
\beq
H^i(t,x^1,x^2+2\pi R  ) = H^i(t,x^1,x^2) e^{i 2\pi R  M}.
\eeq 
If we ignore the infinite tower of the Kaluza-Klein (KK) 
modes, we find the hypermultiplet scalar $\hat H^i(x^1)$ 
of 1+1 dimensions in the lightest mode as
\beq
H^i(t,x^1,x^2) 
= \sqrt{\frac{1}{2\pi R}} \hat H^i(t,x^1) e^{iMx^2}.
\label{presc-SS}
\eeq 
Furthermore, the lightest (constant) mode of 
the gauge field $W_2$ can be identified with the adjoint
scalar $\hat\Sigma$ of the vector multiplet in 1+1 
dimensions
\beq
W_2(t,x^1,x^2)  \equiv -\hat\Sigma(t,x^1). 
\label{w_sigma}
\eeq
We thus have obtained the 1+1 dimensional Lagrangian 
(\ref{eq:waction}) in Sec.~\ref{sub:wall} from the 2+1 
dimensional Lagrangian (\ref{eq:vaction}) 
in Sec.~\ref{sub:vortex} via the SS dimensional reduction.
Both the gauge coupling $\hat g$ and the FI parameter 
$\hat c$ in 1+1 dimensions are also related to the 
corresponding $g$ and $c$ in 2+1 dimensions as 
\beq
\frac{1}{\hat g^2} = \frac{2\pi R}{g^2},\quad
\hat c = 2\pi R c.
\label{paramet}
\eeq

By using the above relation (\ref{presc-SS}), (\ref{w_sigma}) 
and (\ref{paramet}), one can easily 
verify that the BPS equations (\ref{eq:BPS1}) for the vortices 
reduce to 
those (\ref{eq:BPS2}) for the domain walls.
The topological charge (\ref{top:vortex}) of the vortices 
also reduces to that of the walls in Eq.(\ref{top:wall}), 
if integrated only over the fundamental region 
$0 \le x^2 < 2\pi R$.
In addition, there is also a relation between the moduli 
matrix $H_0(z)$ for vortices in Eq.(\ref{sol-vol}) and 
$\hat H_0$ for walls in Eq.(\ref{sol-wall}).
Taking the relation (\ref{presc-SS}) into account, we 
obtain the relation between the moduli matrices
\beq
 H_0(z)\Bigl|_{\rm wall} = \hat H_0 e^{Mz} . 
\label{eq:moduli_relation}
\eeq
The meaning of this equation is as follows; 
When the moduli matrix $H_0(z)$ for {\it vortices} 
has the $z$-dependence as in the right hand side, 
the configurations, 
generated by $H_0(z)$ through 
Eqs.~(\ref{sol-vol}) for {\it vortex} solutions, 
become configurations of {\it domain walls}. 
If we restrict the $V$-equivalence relation (\ref{V-equiv-vol}) 
for vortices to preserve the form (\ref{eq:moduli_relation}), 
we obtain the $\hat V$-equivalence relation (\ref{V-equiv-wall}) 
for domain walls. 
Therefore, the total moduli space of the vortex 
${\cal M}_{\rm v}$ in Eq.(\ref{ms:vortex}) reduces
to that of the domain wall ${\cal M}_{\rm w}$ in 
Eq.(\ref{ms:wall}).

\section{Vortices on Cylinder}\label{sec:v2w}
\subsection{Periodically Arranged Vortices}
\label{subsec:periodic}

\begin{figure}[t]
\begin{center}
\begin{tabular}{ccc}
  \includegraphics[width=40mm]{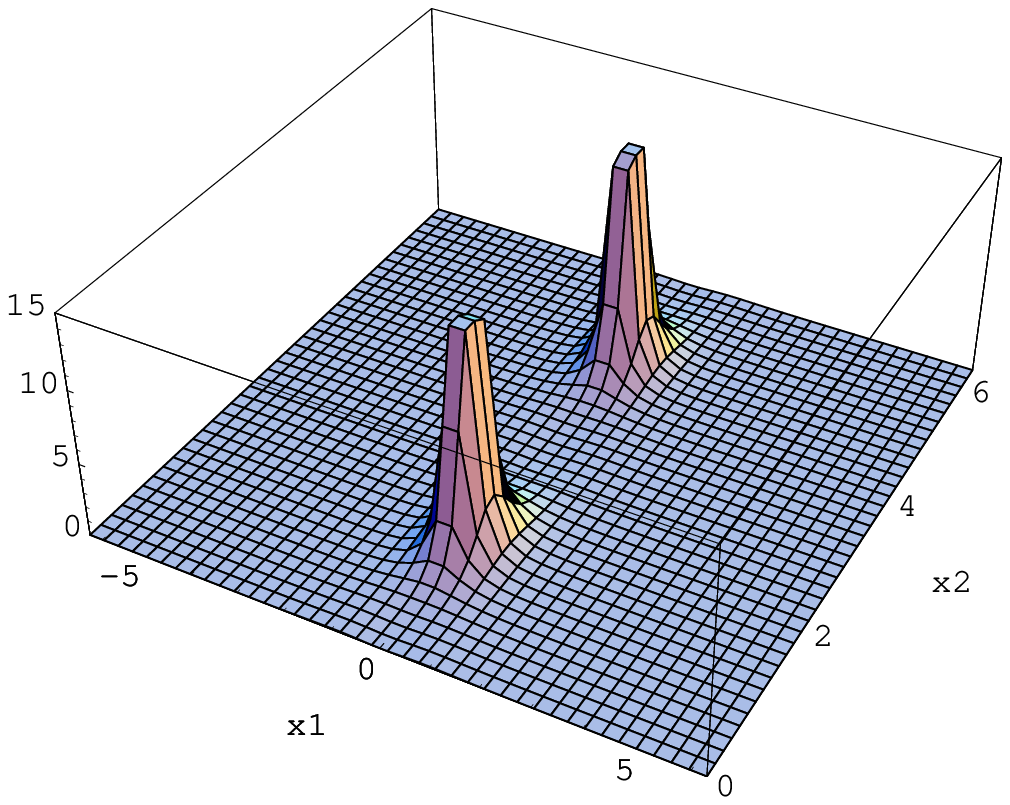}&
  \includegraphics[width=40mm]{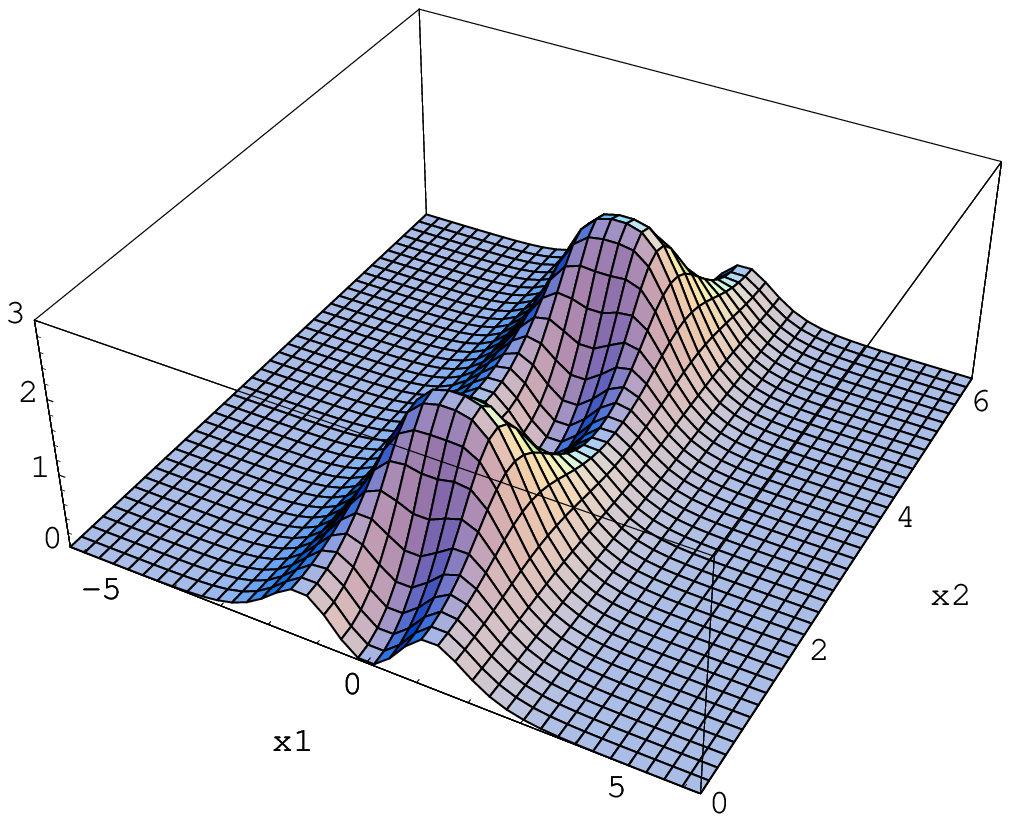}&
  \includegraphics[width=40mm]{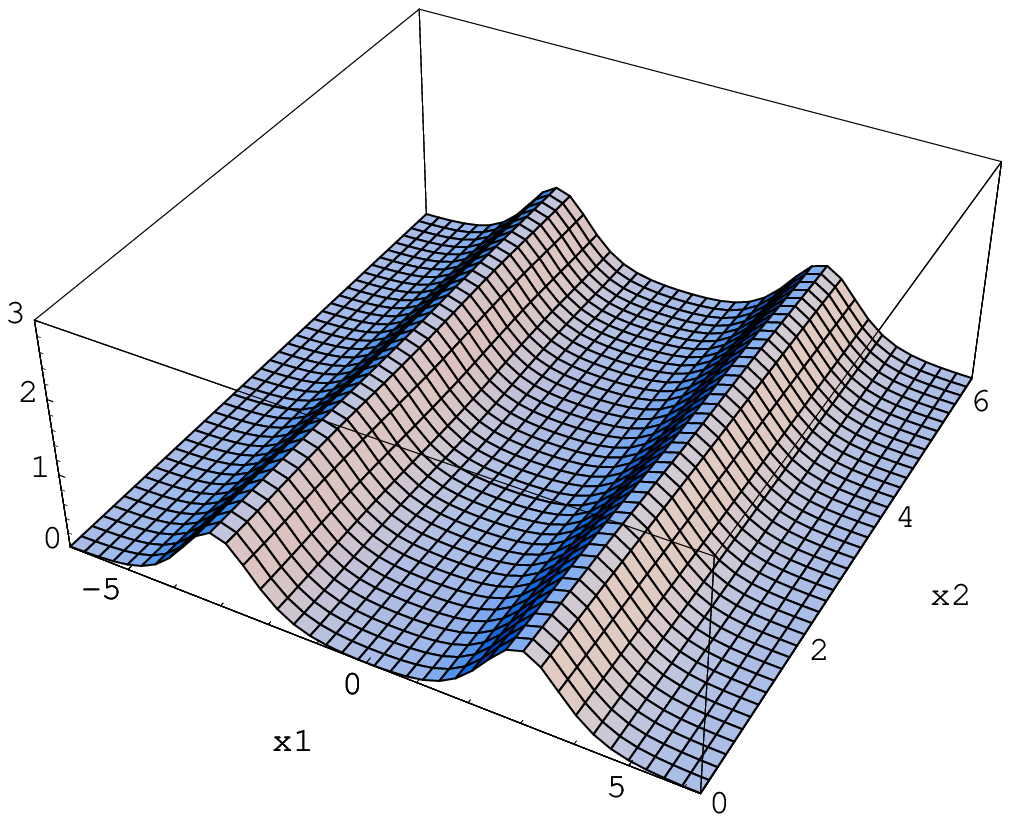} \\
 ${\rm Re} (z_+ - z_-)/R < 0$ & ${\rm Re} (z_+ - z_-) /R \approx 0$ &$ {\rm Re}(z_+ - z_-)/R > 0$
\end{tabular}
\end{center}
\caption{Energy density of the periodically arranged vortices in the strong coupling limit $g \rightarrow \infty$. }
\label{fig:profile}
\end{figure}

\setcounter{equation}{0}
In this section, we construct vortices on $\mathbf R \times S^1$ 
by arranging vortices periodically along the $x^2$-axis and show 
that these vortices can split into two wall-like objects as we vary the the moduli parameters. 
The moduli matrix for periodically arranged vortices is naively given by
\beq
H_0 \sim \left(  a, \prod_{n=-\infty}^{\infty} (z-z_0+i\pi n R )\right ) ,
\label{vor-az0}
\eeq
with $a$ being a complex parameter.
This moduli matrix corresponds to the configuration 
in which there exist two vortices in the strip $0<x^2< 2\pi R $. 
This is the simplest configuration of the periodically 
arranged vortices which can be understood as the configuration 
of the vortices on the cylinder. If we put one vortex in each strip, 
the configuration will not satisfy the periodic boundary condition $H(x^1,x^2) = H(x^1,x^2+2\pi  R)$. By using an appropriate regularization method, this moduli matrix can be rewritten as
\beq
H_0 = \sqrt{c} \left( \hs{1}a,\ R \sinh\left( \frac{z-z_0}{R}\right) \hs{2} \right)  \sim \sqrt{c} \left(\hs{1} 1,\ \frac{R}{a}\sinh\left(\frac{z-z_0}{R}\right) \hs{2}\right) ,\label{eq:sinh}
\eeq
with $a$ being a complex parameter. 
In order to see the transition from vortices to wall-like object, 
it is useful to rewrite this moduli matrix as 
\beq
 H_0 = \sqrt{c} \bpm1,& e^{(z-z_+)/R} - e^{-(z-z_-)/R}\epm.
\eeq
where new moduli parameters $z_+, z_-$ are defined as 
$e^{- z_+/R} \equiv (R/2a) e^{- z_0/R}, 
\hs{2} e^{ z_-/R} \equiv (R/2a) e^{ z_0/R}$. 
Namely, 
$(a,z_0)$ and $(z_+, z_-)$ are good coordinates of moduli 
 for the vortex and the wall-like object respectively to 
extract physics. 
The combination of moduli parameters 
\beq
 {\rm Re} \left(\frac{z_+ - z_-}{R}\right) 
= 2{\rm Re}\left(\log \frac{2a}{R}\right)
\eeq
especially gives the relation between the total size of 
semi-local vortices and distance for two wall-like objects. 
The energy density profiles for 
${\rm Re}(z_+ - z_-)/R < 0$, 
${\rm Re}(z_+-z_-)/R \approx 0$ and 
${\rm Re}(z_+-z_-)/R > 0$ in the strong coupling limit 
are shown in Fig.~\ref{fig:profile}. 
In the strong coupling limit, the theory reduces to 
hyper-K\"ahler nonlinear sigma model on the Higgs branch 
as its target space. 
For infinite gauge coupling, the master equation 
(\ref{eq:master}) becomes a simple algebraic equation 
and can be solved as $ \Omega = \Omega_0 $. 
Then we obtain the exact solutions to the BPS equations 
for vortices (\ref{eq:BPS1}). 
For ${\rm Re}(z_+ - z_-)/R < 0$ ($2|a| < R$), the energy 
density profile is just the same as that of two vortices. 
The sizes of the vortices become larger as we change the 
moduli parameter ${\rm Re}(z_+ - z_-)/R$ to a larger value.  
For ${\rm Re}\hs{1}(z_+ - z_-) /R> 0 $, the vortices 
split into two wall-like objects, whose positions are 
given by ${\rm Re}\hs{1} z_+ $ and ${\rm Re} \hs{1}z_- $. 
Thus we found that the vortices on $ \mathbf R \times S^1$ 
can be split into the domain wall-like objects by changing 
the total scale moduli $a$. 

In the case of $\hs{1}{\rm Re}(z_+ - z_-)/R > 0$, 
the vacua with $H=\sqrt{c}(0,1), \sqrt{c}(1,0), \sqrt{c}(0,1)$ 
 appear in the regions 
$x^1\approx -\infty$, 
$x^1\approx {\rm Re} z_0= {\rm Re}(z_+ + z_-)/2$, 
$x^1\approx \infty$, 
respectively, and the configurations of large size 
($2|a| > R$) vortices on $\mathbf R \times S^1$ look 
like a pair of the wall connecting between the vacua 
$H=\sqrt{c}(0,1), \sqrt{c}(1,0)$ and 
the anti-wall (not anti BPS) connecting between 
$H=\sqrt{c}(1,0), \sqrt{c}(0,1)$. 
An intuitive interpretation of this transition from 
vortices to walls is as follows: Let us put 
two vortices 
on $\mathbf R \times S^1$ and let the total size of 
vortices larger than the radius $2 \pi R$, the vortices 
overlap with vortices living in the next fundamental 
region. 
Then the overlapping vortices form wall-like objects 
with a strip of a different vacuum, 
instead of the ring with a hole for (almost) coincident 
vortices which was observed in Fig.~\ref{hole} and 
Eq.(\ref{eq:hole_coincident_vortices}) in 
Sec.\ref{sub:vortex}. 
In the next section, we will introduce the twisted 
boundary condition in this system.
Then we will obtain deeper understanding of this 
transition.

\subsection{BPS States on Cylinder}\label{subsec:bec}
In this subsection, we explore the 1/2 BPS states on 
$\mathbf R \times S^1$. 
The action is the same as Sec.\ref{eq:vaction} and 
the only difference here is the twisted boundary condition:
\beq
&H^i(x^1,x^2+2\pi R ) = H^i(x^1,x^2) e^{i 2\pi R M}& \\
&M \equiv {\rm diag}(m_1,m_2, \cdots, m_{\NF})&. \nonumber
\eeq
With this boundary condition, we obtain the 1+1 dimensional 
theory which admits domain walls if we perform the SS 
dimensional reduction as we have explained in 
Sec.\ref{sub:ss}. 
However in this subsection, we take into account all the 
KK modes. 
Due to the twisted boundary condition, the vacuum 
conditions are changed into the following equations:
\beq
H^1 (H^1)^\dagger - H^2(H^2)^\dagger 
= c {\mathbf 1}_{\NC}, \hs{5} H^1(H^2)^\dagger =0,  
\nonumber \\
\Sigma^I H^i =0, 
\hs{5} [\Sigma^I,\Sigma^J]= 0, \hs{5} \mathcal D_2 H^i = 0.
\hs{8}
\eeq
Compared with the vacuum conditions (\ref{eq:vacuum1}) 
in Sec.\ref{sub:vortex}, the last equation is added 
since the hypermultiplet scalars $H^i$ cannot be 
constant due to the twisted boundary condition. 
Then the vacua is given by
\beq
 H^{1rA} 
= \sqrt{c}\hs{1} e^{im_{A_r}x^2} \delta^A{}_{A_r} , 
\hs{4}  
H^2=\Sigma^I=0, 
\nonumber \\  
W_2 = -{\rm diag}(m_{A_1},m_{A_2},\cdots,m_{A_{\NC}}). 
\hs{5}
\label{eq:vacuaoncylinder}
\eeq

Therefore, the vacua are discrete and labeled by 
$\langle A_1,A_2,\cdots,A_{\NC}\rangle$ as in the case 
of domain walls. Note that the configurations 
\beq
&H^{1rA} = \sqrt{c}\hs{1} e^{i(m_{A_r}
+n _{A_r}/ R)x^2} \delta^A{}_{A_r},\hs{4}n_{A_r} 
\in \mathbf Z, \hs{4} H^2=\Sigma^I=0, 
&\nonumber \\  &
W_2 = -{\rm diag}(m_{A_1}+n_{A_1}/ R,
\hs{2}m_{A_2}+n_{A_2}/ R,\cdots,
\hs{2}m_{A_{\NC}}+n_{A_{\NC}}/ R) &
\eeq
are also vacuum configurations, which are gauge 
equivalent to the vacua Eq. (\ref{eq:vacuaoncylinder}) 
and related by gauge transformations 
$\Lambda^r{}_s=\delta^r{}_s e^{-i(n_{A_r}/R)x^2}$ 
which are not connected to the unit element of the 
gauge group on $\mathbf R \times S^1$.\\
The 1/2 BPS equations are the same as in 
Sec.\ref{sub:vortex}. 
Since the moduli matrix $H_0(z)$ must satisfy the twisted 
boundary condition, the components of the moduli 
matrix can be written as
\beq
H_0^{rA}(z) 
= e^{m_{A}z} \sum_{n= -\infty}^{\infty} a^{rA}_n e^{nz/R} 
= u^{m_{A} R} \sum_{n= -\infty}^{\infty} a^{rA}_n u^n, 
\hs{5} u \equiv e^{z/R},
\label{H_0-pol}
\eeq
namely polynomials of $u$ and $u^{-1}$ with the factors 
$u^{m_{A} R}$.
 For a configuration which interpolates two vacua 
$\langle A_1,A_2,\cdots,A_{\NC}\rangle$ and 
$\langle B_1,B_2,\cdots,B_{\NC}\rangle$, the topological 
charge $\tilde k$ is given by
\beq
\tilde k &=& \frac{1}{2\pi}\int_{-\infty}^{\infty} 
dx^1 \int_{0}^{2\pi R} dx^2 {\rm Tr} F_{12}  
\nonumber \\ 
&=& \frac{1}{4\pi} \left( \int_{0}^{2\pi R} dx^2 \p_1 
\log \det \Omega_0\Big |_{x^1=\infty} -\int_{0}^{2\pi R} 
dx^2 \p_1 \log \det \Omega_0\Big|_{x^1=-\infty} \right) 
\nonumber \\
&=& k + R\Bigl( \sum_{r=1}^{\NC} m_{A_r} 
- \sum_{r=1}^{\NC} m_{B_r}\Bigl), \hs{10} k \in {\mathbf Z}.
\label{eq:frac_top_charge}
\eeq 
Therefore the highest power of $|u|^2$ minus the lowest 
power of $|u|^2$ in $\det \Omega_0$ gives the topological 
charge. 
The term $R\Bigl( \sum_{r=1}^{\NC} m_{A_r} 
- \sum_{r=1}^{\NC} m_{B_r}\Bigl)$ 
appears because of the twisted boundary condition and 
gives the fractional topological charge. 
We will see later that the integer $k$ gives the number 
of vortices and the fractional topological charge 
corresponds to the tension of domain wall.
Then the moduli space of the solutions with topological 
charge $\tilde{k}$ is expressed as 
\beq
&\mathcal M^{\tilde k}_v \cong \mathcal G / \sim, & \\ 
&\mathcal G \equiv \Bigg \{H_0 \hs{2}\bigg | 
\hs{2}H_0:  \mathbf R \times S^1 
\rightarrow M(\NC \times \NF, \mathbf C),
\hs{2} \overline \p_z H_0 = 0,& \nonumber \hs{10}\\
 &  \hs{25}\det\Omega_0 \xrightarrow[x^1 \to \infty]{} 
|u|^{2l}, 
\hs{2} \det\Omega_0 \xrightarrow[x^1 \to -\infty]{} 
|u|^{2 \tilde l}, 
\hs{2} l-\tilde l = \tilde k \Bigg \}& \nonumber \\ 
& H_0(z) \sim V(z) H_0(z), 
\hs{5} V(z) \in GL(\NC, \mathbf C), 
\hs{5} \overline \p_z V(z) = 0,& \nonumber
\eeq
where again note that all fields appearing in the 
above equations have to satisfy 
the twisted boundary condition.\\
Next we will give the useful estimation method for the 
$x^1$-positions of 
vortices and walls. 
By using this method, we can investigate the transition 
between walls and vortices in general.
The energy density can be roughly estimated as follows:
\beq
{\cal E}
&=&c \hs{1} \partial_z \bar{\partial}_z \log\det\Omega 
\simeq c \hs{1} \partial_z \bar{\partial}_z \log\det\Omega_0
\nonumber 
\eeq
\beq
\det \Omega_0  
&=& \det\Bigl(\frac{1}{c} H_0 H_0^\dagger \Bigl) 
= \sum_{\langle A \rangle} 
\Bigl| \det \frac{1}{\sqrt{c}}H_{0\langle A \rangle} \Bigl|^2  
\nonumber
\eeq
where $\langle A \rangle \equiv \langle A_1\cdots 
A_{\rm N_{\rm C}} \rangle$ 
is the $N_{\rm C}$ set of flavors and the 
$(r,s)$ component of $N_{\rm C}\times N_{\rm C}$ 
matrix $H_{0\langle A \rangle}$ is given by $(H_{0})^{rA_s}$.
Now, it is useful to introduce linear holomorphic functions 
$f_{\langle A \rangle}^n(z)$ as
\beq
\det \left(\frac{1}{\sqrt{c}}H_{0\langle A \rangle} \right)
=\sum_{n}
e^{f_{\langle A \rangle}^n}, 
\ \ \ \ 
f_{\langle A \rangle}^n 
\equiv \Bigl(\frac{n}{R} + \sum_{r=1}^{N_C}m_{A_r} \Bigl) z
 + s_{{\langle A \rangle}} 
\nonumber
\eeq
where the $ s_{{\langle A \rangle}}$ are some functions 
of moduli parameters. 
If there is the region where only one of 
$e^{f_{\langle A \rangle}^n}$ is dominant in 
$\det \Omega_0 $, the energy density vanish there because 
$\p \bar \p \log |\sum e^{f_{\langle A \rangle}^n}|^2=0$.
In contrast, if there is the region where the two of 
$e^{f_{\langle A \rangle}^n}$ 
are comparable, the energy density is non-vanishing. 
Therefore, there 
exist walls or vortices in the region where 
${\rm Re} f_{\langle A \rangle}^n \approx 
{\rm Re}f_{\langle B \rangle}^m$. 
Since the functions $f_{\langle A \rangle}^n$ are linear 
in $z$, the energy density at the transition point 
${\rm Re}f_{\langle A \rangle}^n 
= {\rm Re}f_{\langle B \rangle}^m \ (A\not=B)$ 
is independent of $x^2$, so that there is the wall 
$\langle A\to B\rangle $ at this point.
On the other hand, there are $k$-vortices at the transition 
point 
${\rm Re}f_{\langle A \rangle}^n 
= {\rm Re}f_{\langle A \rangle}^m \ (n-m=k)$.
\\

\begin{figure}[tbp]
\begin{center}
\begin{tabular}{cc}
\includegraphics[width=55mm]{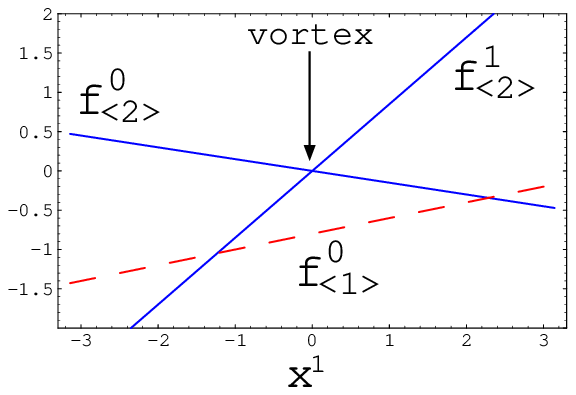} 
& \includegraphics[width=55mm]{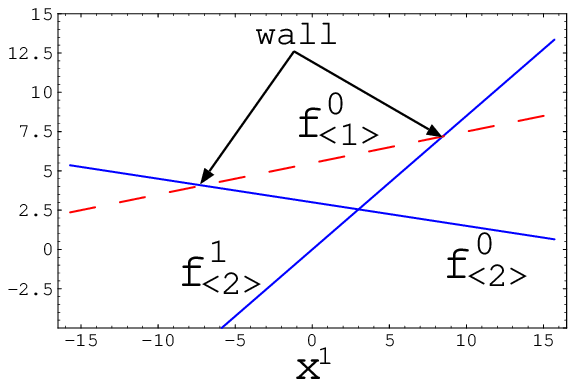}\\(a) vortex 
& (b)two walls  \\ \\
\multicolumn{2}{c}{({\rm i}) Comparison of the 
functions $f_A^n$}\\ \\
\includegraphics[width=55mm]{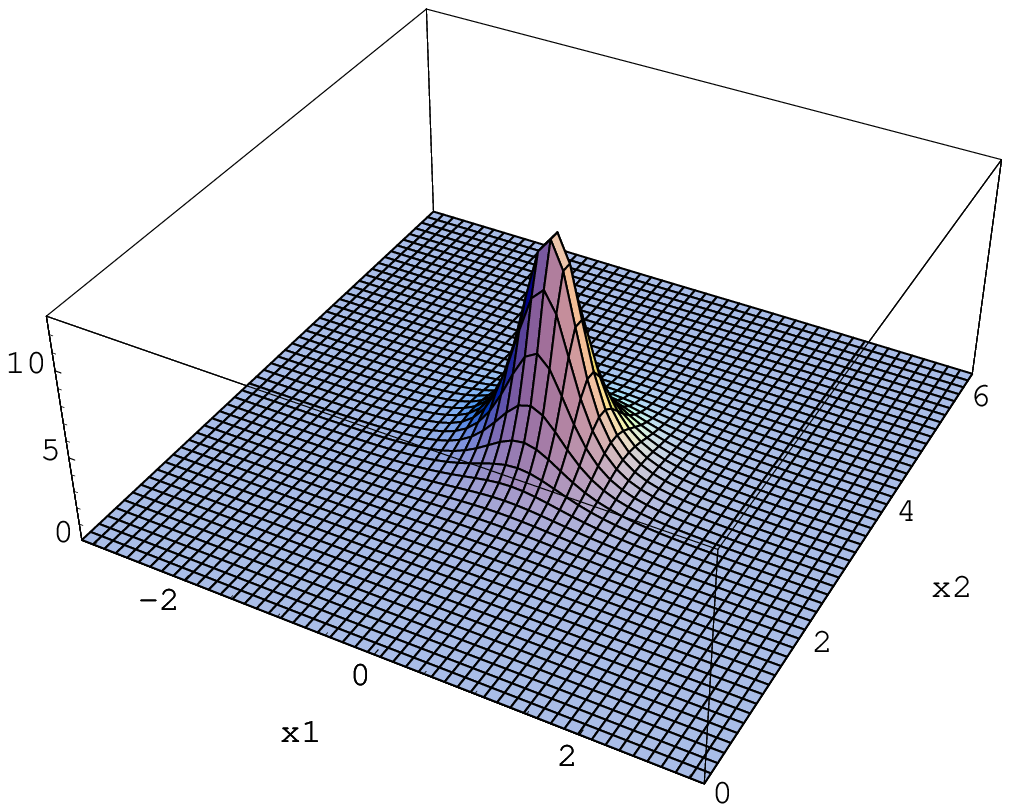} 
& \includegraphics[width=55mm]{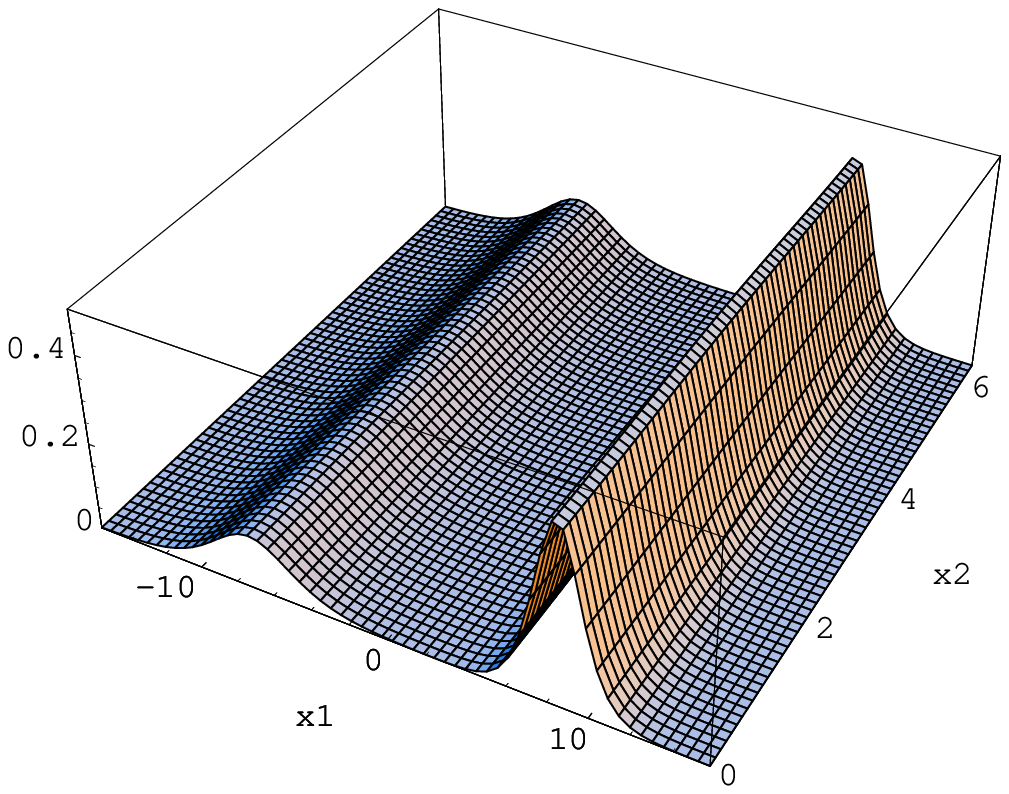} \\(a)&(b)\\ \\
\multicolumn{2}{c}{({\rm ii}) Energy density} \\ \\ 
\includegraphics[width=55mm]{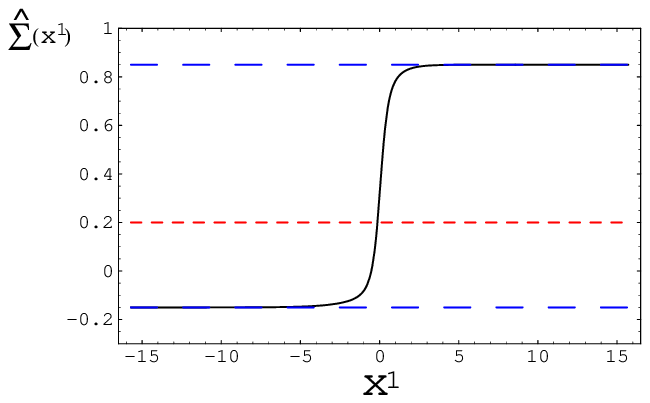} 
& \includegraphics[width=55mm]{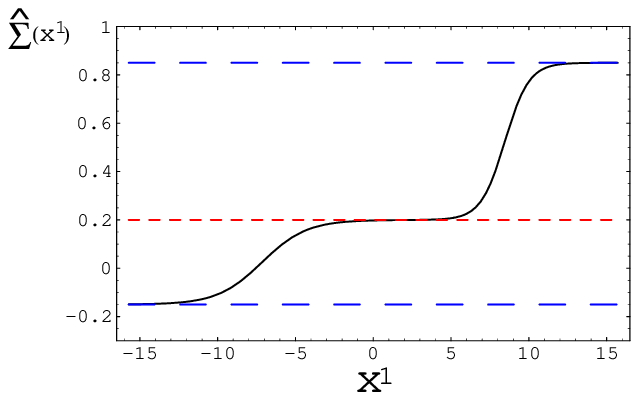} \\
(a)&(b)\\ \\
\multicolumn{2}{c}{({\rm iii}) vector multiplet 
scalar $\Sigma(x^1)$}
\end{tabular}
\end{center}
\caption{Examples: $R=1, \hs{2} m_1=0.2,
\hs{2} m_2=-0.15$ \hs{4} {}  (a) $a_1=e^{-0.6},a_2=1$, 
(b) $a_1=e^{5.5},a_2=e^{3}$}
\label{fig:example}
\end{figure}

We now see some explicit examples in the case of 
$\NC=1,\hs{2} \NF =2$. 
First, we consider $k=0$ case. 
The most general moduli matrix is given by
\beq
H_0 = \sqrt{c} \bpm 1, & a \epm e^{Mz} 
= \sqrt{c}\bpm 1, & a  \epm \bpm u^{m_1 R} & 0 
\\ 0 & u^{m_2 R} \epm,
\eeq
where $a$ is a complex parameter. 
For this moduli matrix, there are no contributions 
from the KK modes to any physical quantities. 
Therefore this moduli matrix corresponds to the domain 
wall configuration as stated in Sec.\ref{sub:ss}.
\\
Next, we investigate $\tilde{k}=1$ case. 
This is the simplest case where 
KK-modes contribute to the topological charge. 
The moduli matrix is given by
\beq
&H_0 = \sqrt{c} \bpm a_1,& e^{z/R}+a_2 \epm e^{Mz} 
\sim \sqrt{c} \bpm 1, 
& \frac{1}{a_1}(u + a_2) \epm \bpm u^{m_1 R} 
& 0 \\ 0 & u^{m_2 R} \epm& 
\label{eq:moduli}
\\ 
&a_1 \in \mathbf C, \hs{5} a_2 \in \mathbf C-\{0\}.& 
\nonumber
\eeq
Note that in the strong coupling limit, the master 
equation (\ref{eq:master}) can be solved as 
$\Omega = \Omega_0 = \frac{1}{c} H_0 H_0^{\dagger}$.
 Therefore, the energy density for this moduli matrix 
in the strong coupling limit can be expressed as
\beq
\mathcal E(x^1,x^2) 
 &=& c \hs{1} \p_z \bar \p_z \log 
 \left( |e^{f_{\langle 1\rangle} ^0}|^2 
 + |e^{f_{\langle 2\rangle}^1}
+e^{f_{\langle 2\rangle}^0}|^2 \right), 
\label{eq:energydensity}
\eeq
where we have defined the functions 
$f_{\langle A\rangle}^n$, 
which are linear in $z$, by
\beq
f^0_{\langle 1\rangle }(z) = m_1 z  , 
\hs{35}\nonumber \\ 
f^0_{\langle 2\rangle }(z)= m_2 z 
+ \log \frac{a_2}{a_1},\hs{5}  
f^1_{\langle 2\rangle }(z)
= \left(\frac{1}{R}+m_2 \right) z -\log a_1.
\label{def-f_A^n}
\eeq 
Now we will explain the fact that one vortex can be 
decomposed into two walls by changing the scale moduli.
We begin with the region in the moduli space where 
the profiles of the functions $f_{\langle A \rangle}^n$ 
are like Fig.~\ref{fig:example}(i)-(a). 
We can see that there exist one vortex at the transition 
point where 
$f_{\langle 2 \rangle}^0= f_{\langle 2 \rangle}^1$. 
Next, let us change the scale moduli $a_1$ to a larger 
value and consider the case where the profiles of the 
functions $f_{\langle A \rangle}^n$ are like 
Fig.~\ref{fig:example}(i)-(b). 
In this region of moduli space, two walls appear at 
two transition points; 
$f_{\langle 2 \rangle}^0= f_{\langle 1 \rangle}^0 $ and 
$f_{\langle 1 \rangle}^0 = f_{\langle 2 \rangle}^1$.
The corresponding energy density are shown in 
Fig.~\ref{fig:example}(ii)-(a),(b).
Therefore we can conclude that one vortex can be 
decomposed into 
the wall $\langle 0\to 1 \rangle$ and the anti 
wall $\langle 1\to 0 \rangle$
by changing the scale moduli parameter. 

Now we return to Sec.\ref{subsec:periodic}  
and comment on the relation between the twisted boundary 
condition and the periodic one. 
If we turn off the the parameters $m_i$, we may well 
deduce that the former reduces to the latter. 
In this limit, however, we have to take into account 
the periodic boundary condition 
because that some of $f_{\langle A \rangle}^n$ belonging 
to the same flavor reduce to have the same slope. 
As a result, we need at least two vortices at the same $x^1$ 
point in the fundamental region.
Except for this technical point, whole argument are 
essentially the same as in the twisted one.
 For ${\rm Re}(z_+-z_-)<0$, since the configuration 
is the same vacuum configuration in the region where 
${\rm Re}\hs{1}z \ll {\rm Re}z_0={\rm Re}(z_++z_-)/2$ and 
${\rm Re}\hs{1}z\gg {\rm Re}z_0$, the vortices 
appear at ${\rm Re}\hs{1}z={\rm Re}z_0$. 
For ${\rm Re} (z_+-z_-)>0$, the configuration approaches 
the vacuum $H=\sqrt{c}(0,1)$ at 
${\rm Re} \hs{1} z \rightarrow \pm \infty$ and 
$H \approx \sqrt{c}(1,0)$ in the region where 
${\rm Re}\hs{1}z_- < {\rm Re}\hs{1} z < {\rm Re} \hs{1}z_+$. 
Since there are two transition points of the different 
vacua, two walls appear at $z=z_+$ and $z=z_-$.

Next we calculate the Wilson loops around the $S^1$ of 
the cylinder and relate this to $\hat \Sigma(x^1)$ by
\beq
\hat \Sigma (x^1) \equiv -\frac{1}{2\pi R}\log\Bigl[ \mathbf{P} \ 
{\rm exp}\int_0^{2\pi R}dx^2 W_2 (x_1, x_2) \Bigl].
\label{eq:defsigma}
\eeq
 {
If the gauge field $W_2$ is independent of $x^2$, 
the definition of $\hat \Sigma(x^1)$ in Eq.~(\ref{eq:defsigma}) 
reduces to the one in Eq.~(\ref{w_sigma}). 
Therefore 
this quantity corresponds to the scalar field 
in the vector multiplet after the dimensional reduction.
In general $\hat \Sigma (x^1)$ exhibits kinks as seen below. 
To illustrate this we consider 
the case of $\NC = 1,\hs{2} \NF = 2$ and $\tilde k=1$. 
We take the strong gauge coupling limit 
$ g \rightarrow \infty$ for simplicity. 
Then the gauge field can be obtained explicitly as 
}
\beq 
W_{2} = - S^{-1} \p_1 S, \hs{5} S 
= \sqrt{H_0 H_0^{\dagger} \over c},
\eeq
with $H_0$ given in Eq.~(\ref{eq:moduli}). 
Then, we obtain $\hat \Sigma(x^1)$ by integrating the 
gauge field around $S^1$:
\beq
\hat \Sigma(x^1) &=& 
- \frac{1}{2\pi R} \int_0^{2\pi R} dx^2 W_2(x^1, x^2) 
 \label{eq:Sigma} \\
&=& \frac{|a_1|^2(m_1-m_2-\frac{1}{2R})e^{2m_1x^1}
+\frac{1}{2R}(e^{2x^1/R}-|a_2|^2 )e^{2m_2 x^1}}
{\sqrt{(e^{2(m_2+1/R)x^1}+|a_1|^2e^{2m_1x^1}
+|a_2|^2e^{2m_2x^1})^2-4|a_2|^2 e^{2(2 m_2+1/R)x^1}}} 
\nonumber \\&& +m_2+\frac{1}{2R} \nonumber,
\eeq
Fig.~\ref{fig:example}({\rm iii}) (a),(b) show 
examples of 
the vector multiplet scalar $\hat \Sigma(x^1)$. 

 {
Although in the small size limit $|a_1| \rightarrow 0$ 
the semi-local vortex reduces to the ANO vortex
for a finite gauge coupling constant, 
it corresponds to small lump singularity
in the strong coupling limit. 
Correspondingly, 
as shown in the Fig.~\ref{fig:singular}, 
$\hat \Sigma(x^1)$ becomes a step function  
in the small size limit $|a_1| \rightarrow 0$: 
\begin{figure}[h]
\begin{center}
\includegraphics[width= 60mm]{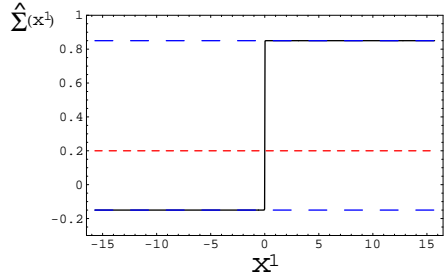}
\end{center}
\caption{Profile of small size limit of $\hat \Sigma(x^1)$ in the strong gauge coupling limit}
\label{fig:singular}
\end{figure}
\beq
\hat \Sigma(x^1) &\rightarrow& \frac{1}{2R} \frac{(e^{2x^1/R}-|a_2|^2)e^{2m_2 x^1}}{\sqrt{((e^{2x^1/R}-|a_2|^2)e^{2m_2 x^1})^2}} +m_2 + \frac{1}{2R}
  \nonumber \\
 &=& \frac{1}{2R} {\rm sign}  \left(x^1 - \frac{R}{2} \log |a_2|^2 \right) 
   + m_2 + \frac{1}{2R} . \label{eq:singular}
\eeq
This result reflects the fact that the size of the vortex is encoded to the size of kink of $\hat \Sigma(x^1)$.
In the next section, we discuss the brane construction 
of our model. 
In the brane construction, our theory corresponds to the 
worldvolume theory of the \rm D3 branes. 
After taking the T-duality transformation, 
$\hat \Sigma(x^1)$ parameterizes the position of the {\rm D}2 
branes in the dual direction. 
Therefore $\hat \Sigma(x^1)$ can be understood as the 
position of {\rm D}-branes in the dual direction. 
}

Next we consider $R \rightarrow 0$ limit and show that 
the finite energy solutions are ordinary domain wall 
solutions after taking the limit. 
When we take the limit, we keep both the FI parameter 
$\hat c$ and the masses $m_A$ in the (1+1)-dimension 
fixed. 
We again use the moduli matrix in Eq.~(\ref{eq:moduli}) 
as an example. 
If $a_1$ is small, the corresponding configuration is 
a one vortex configuration. 
Since the mass of the vortex is 
$2 \pi c = \frac{\hat c}{R}$, the vortex decouples 
in the $R \rightarrow 0$ limit. 
If the parameter $a_1$ is sufficiently large, 
there are two walls with masses 
$2\pi c R (m_1-m_2) = \hat c (m_1-m_2)$ and 
$2 \pi c(1-R(m_1-m_2)) = \hat c(\frac{1}{R}-(m_1-m_2))$. 
The wall with mass $\hat c (m_1-m_2)$ is the ordinary 
domain wall appearing in the (1+1)-dimensional theory. 
The other wall with mass 
$\hat c(\frac{1}{R}-(m_1-m_2))$ 
becomes infinitely massive. 
To obtain a meaningful limit $R\rightarrow 0$, 
we need to let the position of this infinitely massive 
wall to either infinity $x^1 \rightarrow \pm \infty$. 
In this sense, this infinitely massive wall decouples. 
In the case of 
$\tilde k =k+R(m_1-m_2)\geq 1$ for $k=1$, 
an additional domain wall 
with mass $\hat c (m_1-m_2)$ exists. 
However, one should note that the walls are inevitably 
ordered in such a way that the infinitely massive wall 
is sandwiched between this wall and the previously 
mentioned wall with the same mass $\hat c (m_1-m_2)$. 
Therefore either one of the two walls with 
the mass $\hat c (m_1-m_2)$ must go to infinity in the 
$R\rightarrow 0$ limit, since it should be outside of 
the infinitely massive wall which goes to infinity in 
the limit. 
For arbitrary $\NC$ and $\NF$ there are $\NF$ kinds of 
domain walls and one of these walls has mass which will 
be infinite in the $R \rightarrow 0$. 
For a finite $R$, the numbers of the 
$\NF$ types walls are not restricted.
However, as in the case of the previous example, 
the infinitely massive walls take some of the other walls
to the infinity in the $R \rightarrow 0$ limit. 
For a finite energy solution, at most
$(\NF-\NC) \times \NC$ walls are left in the 
$R \rightarrow 0$ limit and these are the ordinary domain walls
 appearing in the (1+1)-dimensional theory.

\section{Brane Configurations and T-duality}\label{sec:brane}
\begin{figure}[h]
 \begin{center}
  \includegraphics[width= 60mm]{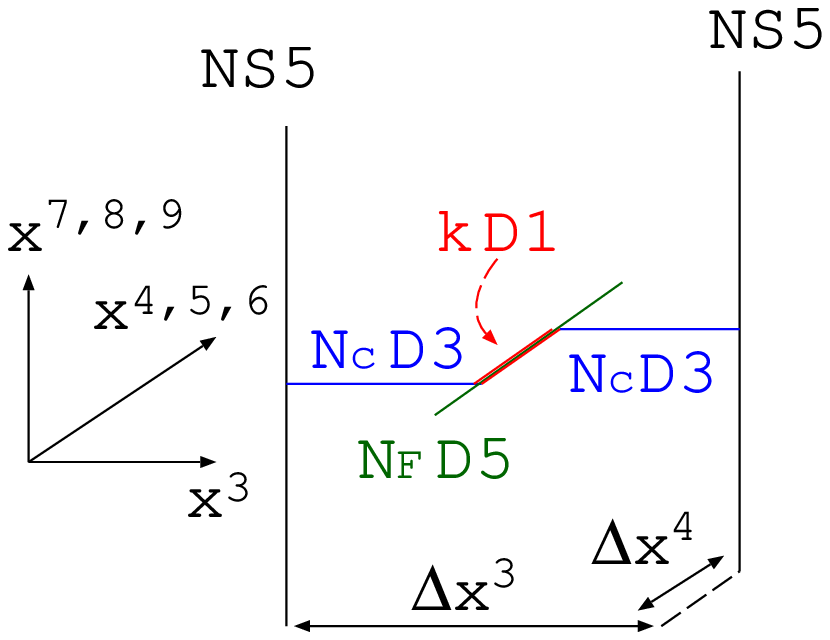}
  \includegraphics[width= 60mm]{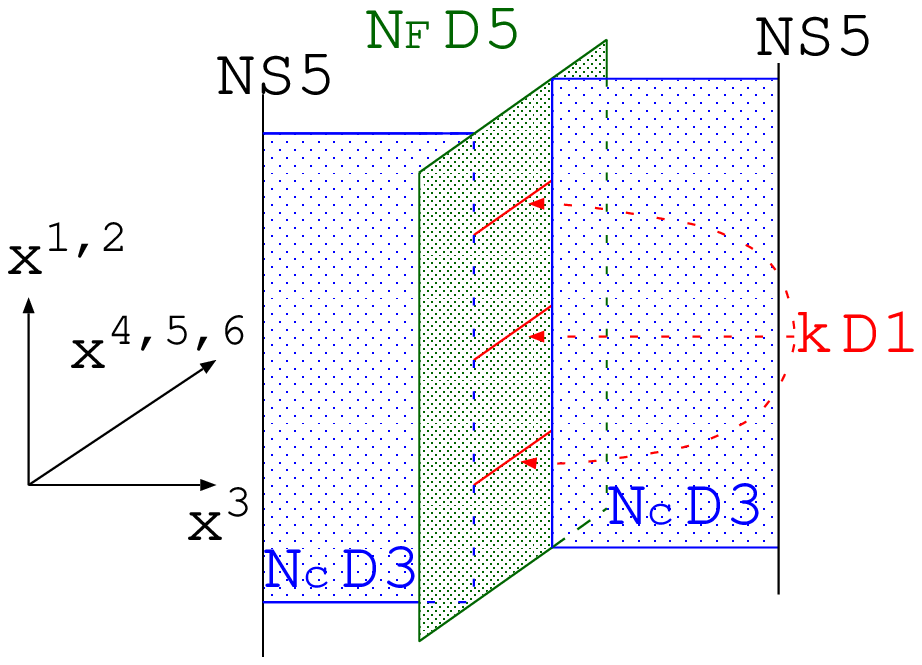}
 \end{center}
 \caption{ Brane configuration for $k$ vortices}
 \label{fig:vortices}
\end{figure}

\setcounter{equation}{0}
In this section, we discuss the relation between vortices and domain walls by using the brane configurations. 
First we consider a brane configuration for 
vortices~\cite{Hanany:2003hp,Hanany:2004ea}. 
Fig.~\ref{fig:vortices} shows the brane configuration for vortices and 
Table \ref{tab:vortices} shows 
the directions in which the branes extend. 
\begin{table}[h]
\begin{center}
\begin{tabular}{!{\tvline}c!{\tvline}cccccccccc!{\tvline}}
\thline
&$x^0$&$x^1$&$x^2$&$x^3$&$x^4$&$x^5$&$x^6$&$x^7$&$x^8$&$x^9$\\
\thline
$\NC$ $\rm D3$&$\circ$&$\circ$&$\circ$&$\circ$&$-$&$-$&$-$&$-$&$-$&$-$ \\
\hline
$\NF$ $\rm D5$&$\circ$&$\circ$&$\circ$&$-$&$\circ$&$\circ$&$\circ$&$-$&$-$&$-$ \\
\hline 
$2\hs{2}{\rm NS}5$&$\circ$&$\circ$&$\circ$&$-$&$-$&$-$&$-$&$\circ$&$\circ$&$\circ$ \\
\hline
$k\hs{2}\rm D1$&$\circ$&$\times$&$\times$&$-$&$\circ$&$-$&$-$&$-$&$-$&$-$ \\
\thline
\end{tabular}
\end{center}
\caption{ Brane configuration for $k$ vortices: Branes are extended along directions denoted by $\circ$, and are not extended
along directions denoted by $-$. The symbol $\times$ denotes the
codimensions of the k D1-branes on the worldvolume of the D3-branes
excluding the $x^3$ which is a finite line segment.}
\label{tab:vortices}
\end{table}
Since the coincident $\NC$ $\rm D3$ branes which are 
stretched between two ${\rm NS}5$ branes have the finite 
length $\Delta x^3$, the theory on the worldvolume of 
the $\rm D3$ branes reduces to $(2+1)$-dimensional 
$U(\NC)$ gauge theory with a gauge coupling 
$\frac{1}{g^2} = |\Delta x^3| \tau_3 l_s^4 
= \frac{|\Delta x^3|}{g^{(B)}_s}$, where $g^{(B)}_s$ is 
the string coupling constant in type IIB string theory
and $\tau_3 = 1/g^{(B)}_s l_s^4$ is the D3-brane tension.
Since the positions of the $\NF$ $\rm D5$ branes 
in the $x^7$-,$x^8$- and $x^9$-directions coincide 
with those of the $\rm D3$ branes, 
the $\NF$ hypermultiplets  
in the D3 brane worldvolume theory 
coming from D3-D5 strings are massless. 
The separations of two ${\rm NS}5$ branes in the 
$x^4$-,$x^5$- and $x^6$-directions correspond to 
the triplet of the FI parameters $c^a$, which we 
choose as $c^a = (0,0,c=\Delta x^4/g^{(B)}_s l_s^2 >0)$, 
where $l_s$ is the string length. 
The symbol $\times$ in Table \ref{tab:vortices} denotes 
the codimensions of the $k\hs{2} \rm D1$ branes on the 
worldvolume of the $\rm D3$ branes. 
Since the $x^3$ direction of the $\rm D3$ brane worldvolume 
is a finite line segment, the $\rm D1$ branes are 
interpreted as codimension-two objects on the $\rm D3$ branes. 
 {
Moreover, the energy $\tau_1 \Delta x^4=\frac{\Delta x^4}{g^{(B)}_s l_s^2}=c$ 
of each $\rm D1$ brane 
corresponds to the energy of a vortex given in Eq.(\ref{top:vortex}).
}
Therefore the $k \hs{2} \rm D1$ branes correspond to $k$ 
vortices in the $\rm D3$ brane worldvolume theory. 

Next, we consider the T-dual picture of this configuration. 
We compactify the $x^2$ direction on $S^1$ with radius $R$ and take T-duality 
along that direction. 
Table \ref{tab:walls} shows 
the directions in which the branes extend
after T-duality transformation. 
\begin{table}[h]
\begin{center}
\begin{tabular}{!{\tvline}c!{\tvline}cccccccccc!{\tvline}}
\thline
&$x^0$&$x^1$&$x^2$&$x^3$&$x^4$&$x^5$&$x^6$&$x^7$&$x^8$&$x^9$\\
\thline
$\NC$ $\rm D2$&$\circ$&$\circ$&$-$&$\circ$&$-$&$-$&$-$&$-$&$-$&$-$ \\
\hline
$\NF$ $\rm D4$&$\circ$&$\circ$&$-$&$-$&$\circ$&$\circ$&$\circ$&$-$&$-$&$-$ \\
\hline 
$2\hs{2}{\rm NS}5$&$\circ$&$\circ$&$\circ$&$-$&$-$&$-$&$-$&$\circ$&$\circ$&$\circ$ \\
\hline
$k\hs{2}\rm D2'$&$\circ$&$\times$&$\circ$&$-$&$\circ$&$-$&$-$&$-$&$-$&$-$ \\
\thline
\end{tabular}
\end{center}
\caption{T-dualized configuration: Branes are extended along directions denoted by $\circ$, and are not extended
along directions denoted by $-$. The symbol $\times$ denotes the
codimensions of the k D2'-branes on the worldvolume of the D2-branes,
excluding the $x^3$ which is a finite line segment.
\label{tab:walls}
}
\end{table}
Before T-dualizing, we turn on a constant background gauge 
field  $A_2 = {\rm diag} (m_1,\cdots,m_{\NF})$ on the $\rm D5$ brane worldvolume, that is a non-trivial Wilson loop around $S^1$ on the $\rm D5$ branes. 
With this Wilson loop, the resulting $\rm D4$ branes 
split in the $x^2$-direction and  {their positions are determined as $X^2_A=2 \pi l_s^2 m_A \hs{2} (A = 1,\cdots,\NF)$. These separations give hypermultiplet masses in the $\rm D2$ brane worldvolume theory. 
The worldvolume theory of the $\rm D2$ branes is 
$(1+1)$-dimensional gauge theory with a gauge coupling 
$\frac{1}{\hat g^2} = \frac{\Delta x^3 l_s}{g^{(A)}_s} = 
\frac{\Delta x^3 R}{g^{(B)}_s}$
and the FI parameter
$\hat c = \frac{\Delta x^4}{g^{(A)}_s l_s} = \frac{\Delta x^4 R}{g^{(B)}_s l^2_s} $, where $g^{(A)}_s$ is the string coupling constant in type IIA string theory.}

\begin{figure}[h]
\begin{center}
\includegraphics[width= 50mm]{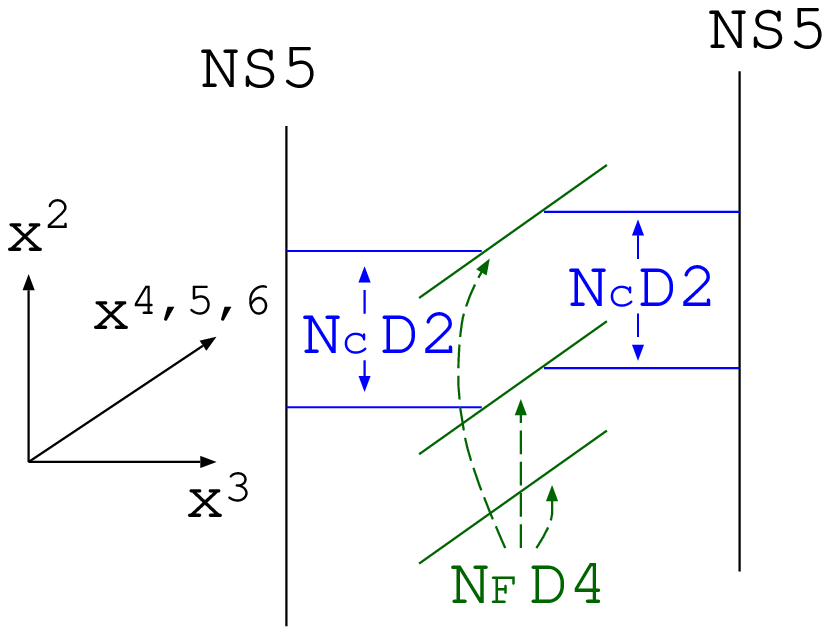}
\end{center}
\caption{vacuum configuration $(k=0)$}
\label{fig:k=0}
\end{figure}
For $k=0$, as shown in Fig.~\ref{fig:k=0}, each $\rm D2$ 
brane ends on one of the $\rm D4$ branes and at most one 
$\rm D2$ brane can be stretched between a $\rm D4$ brane 
and a ${\rm NS5}$ brane due to the s-rule~\cite{HW}. 
These configurations correspond to the vacua in the 
$\rm D2$ brane worldvolume theory. 

\begin{figure}[h]
 \begin{center}
  \includegraphics[width= 70mm]{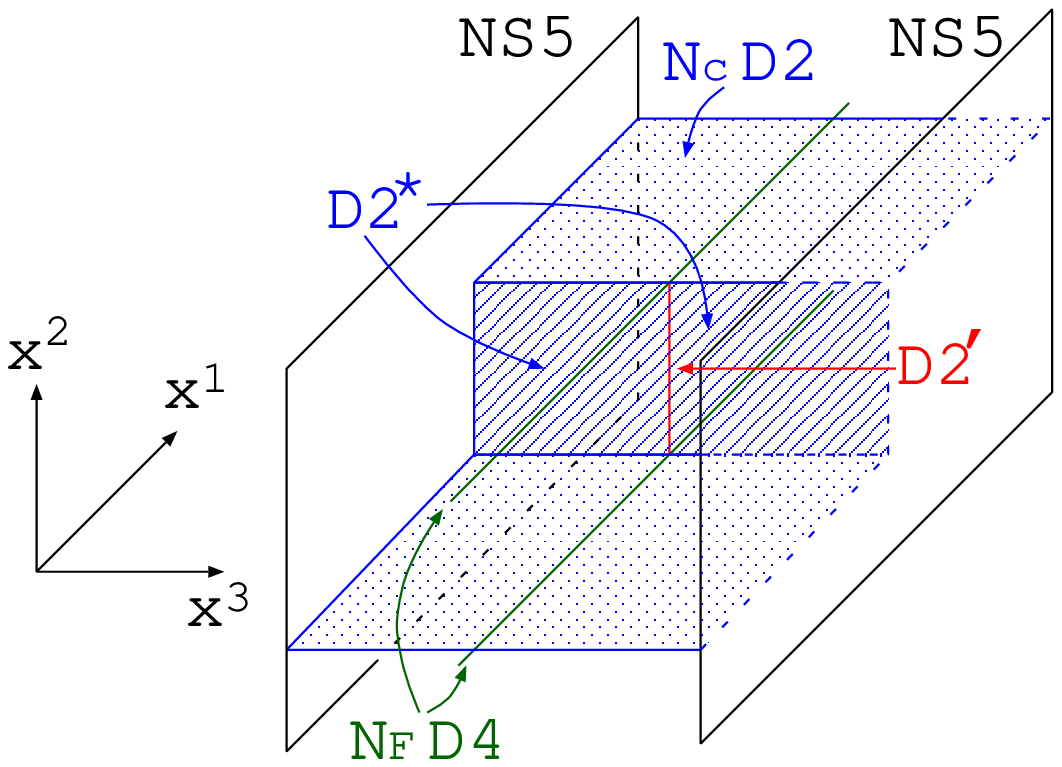}
 \end{center}
 \caption{Brane configuration for a wall}
 \label{fig:wall}
\end{figure}

For $k \not = 0$, the $\rm D1$ branes transform into 
$\rm D2$ branes stretched between the $\rm D4$ branes 
and we denote these $\rm D2$ branes as $\rm D2'$.
\begin{figure}[h]
\begin{center}
\begin{tabular}{cccc}
\includegraphics[width=15mm]{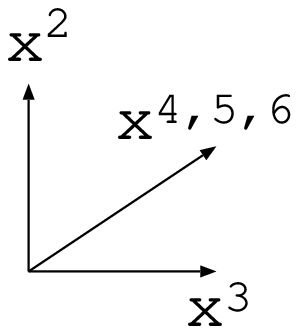} &
\includegraphics[width=35mm]{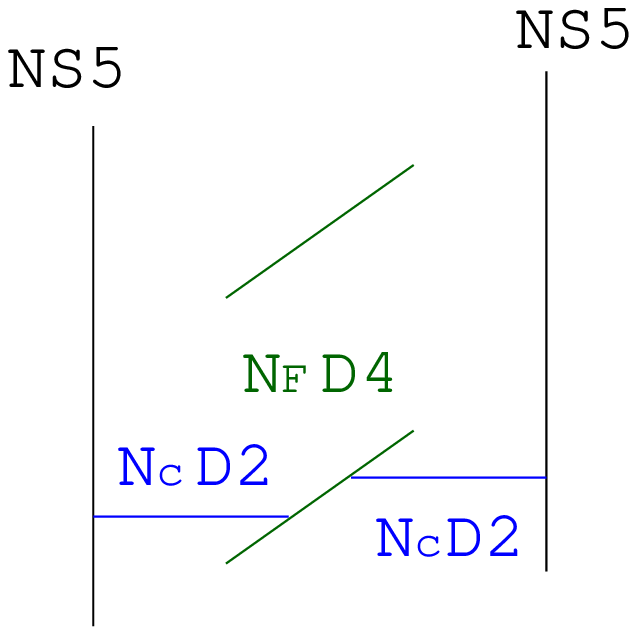}&
\includegraphics[width=35mm]{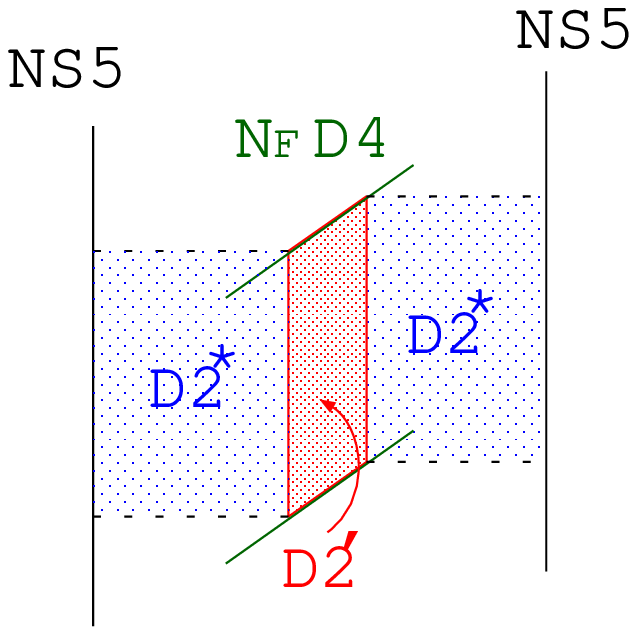} &
\includegraphics[width=35mm]{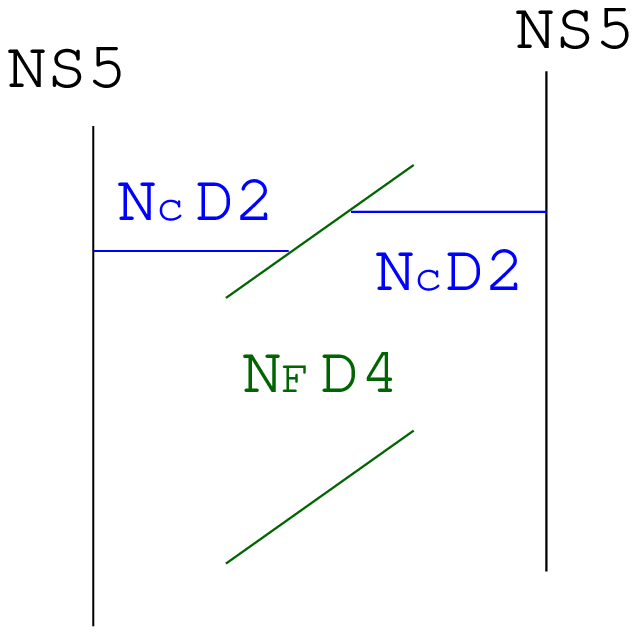} \\
&$x^1 = -\infty$ & $x^1 = x_0$ & $x^1=+\infty$
\end{tabular}
\end{center}
\caption{ T-dualized configurations for $k=1$: 
the $\rm D2'$ brane is assumed to be located at $x^1=x_0$.}
\label{fig:k=1}
\end{figure} 
Fig.~\ref{fig:k=1} shows the brane configurations for $k=1$. 
Since the $\rm D2$ branes are attached to different 
$\rm D4$ branes at $x=-\infty$ and $x=+\infty$, there 
must exist $\rm D2$ branes which connect the $\rm D2$ 
branes ending on different $\rm D4$ branes. 
These $\rm D2$ branes correspond to $\rm D2'$ in Fig.~\ref{fig:k=1} and Fig.~\ref{fig:wall}. 
In addition, there must be $\rm D2$ branes ($\rm D2^\ast$ 
in Fig.~\ref{fig:k=1} and Fig.~\ref{fig:wall}) at a points where the $\rm D2'$ 
branes are located and the $\rm D2'$ branes end on these 
$\rm D2^\ast$ branes. 
Fig.~\ref{fig:wall} shows the resulting brane configuration 
for $k=1$ which corresponds to a BPS domain wall 
configuration in the $\rm D2$ brane theory.
 {
Indeed, we can easily calculate the energy of the domain walls 
from this brane configuration in the strong coupling limit. 
Since the gauge coupling $\frac{1}{\hat g^2}$ is proportional to $\Delta x^3$, 
the $\rm D2^{\ast}$ branes disappear in the limit 
$\hat g \rightarrow \infty$. The remaining $\rm D2'$ branes have the energy $\tau_2 \hs{1} \Delta x^4  \hs{1} l_s^2 \Delta m = \frac{\Delta x^4 \Delta m}{g^{(A)}_s l_s}=\hat c \hs{1} \Delta m$ and this corresponds to the energy (\ref{top:wall}) of a domain wall. 
}

We thus conclude that the vortices and the domain walls are related 
by T-duality. 
This relation is analogous to the relation between instantons and monopoles. 
In Fig.~\ref{fig:vortices} the ${\rm D}1$ branes, 
which correspond to the vortices in the $\rm D3$ brane worldvolume theory, 
can be thought of as instantons in the $\rm D5$ brane worldvolume theory. 
Similarly, the $\rm D2'$ brane in the Fig.~\ref{fig:wall} can be 
thought of as a monopole in the $\rm D4$ brane worldvolume theory. 
For this reason, the relation between vortices and domain walls 
is similar to the relation between instantons and monopoles. 

\begin{figure}[h]
\begin{center}
\begin{tabular}{ccc}
\includegraphics[width=15mm]{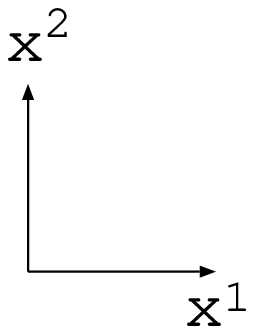} &
\includegraphics[width=55mm]{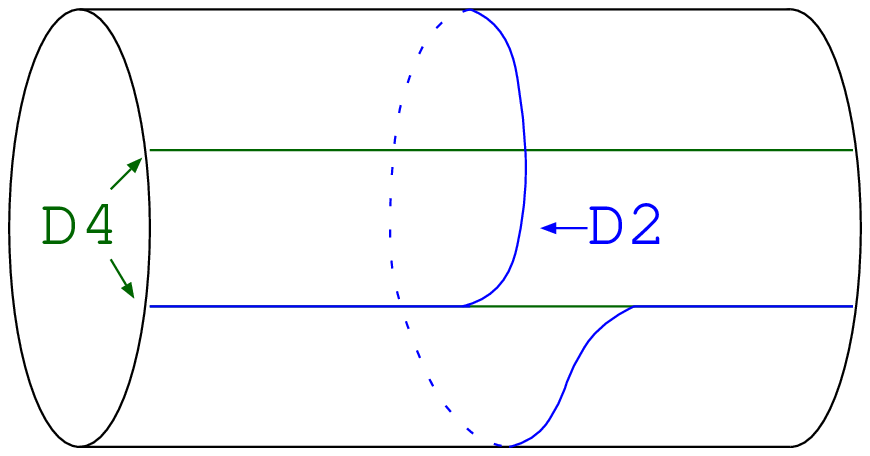} &
\includegraphics[width=55mm]{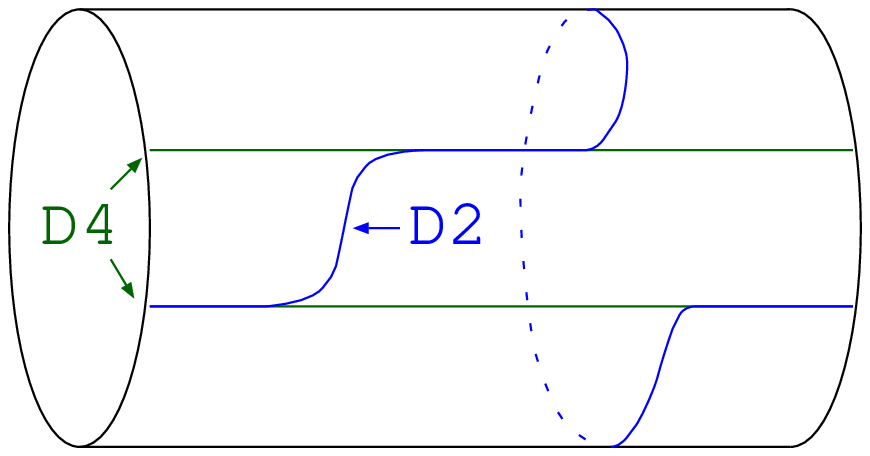} \\
& (a) & (b)
\end{tabular}
\end{center}
\caption{ T-dual picture for $\NF=2, \NC=1, k=1$. \label{fig:tk=1}}
\end{figure}

Before taking T-duality, 
D1-brane end-points on D3-branes (in Fig.~\ref{fig:vortices}-(b)) 
are not actually points with zero size 
from the point of view of the field theory on the D3 branes. 
Instead, they have the vortex size $1/g \sqrt c$ 
(or lager size in the case of semi-local vortices). 
This actual configuration can be read from gauge fields. 
Correspondingly, 
this information is expected to be mapped
into the Wilson loop by taking T-duality. 
In fact, although 
we have naively assumed 
the positions of the $\rm D2$ branes 
in the $x^2$ direction so far in this section, 
the actual positions of the {\rm D}-branes 
must be parameterized by $\hat \Sigma(x^1)$ Eq.~(\ref{eq:defsigma}). 
We have already calculated $\hat \Sigma(x^1)$ in Sec.\ref{subsec:bec} 
for $\NF=2, \NC=1, k=1$ as in Eq.~(\ref{eq:Sigma}), 
and have found that the vortex configuration shown 
in Fig.~\ref{fig:example} (a) [for (i) - (iii)] interpolates the same vacuum. 
In the brane picture, 
the vortex corresponds to the \rm D2 brane winding 
around the $S^1$ of the cylinder 
with exhibiting a kink  
as in Fig.~\ref{fig:tk=1} (a). 
The size of this kink in $x^1$ 
is the ANO vortex size $1/g\sqrt c$ from the construction. 
Note that the scalar field $\hat \Sigma(x^1)$ has period $1/R$.
This vortex can be decomposed into two walls 
by changing the size of the vortex as in Fig.~\ref{fig:example} (b) [for (i) - (iii)]. 
In this configuration, the \rm D2 brane is attached to the same \rm D4 brane at $x^1 \rightarrow \pm \infty$ 
with exhibiting kinks twice as in Fig.~\ref{fig:tk=1} (b).
In the middle region, the \rm D2 brane ends on the other \rm D4 brane. 
Therefore, there are two walls at the points where the \rm D2 brane change 
the \rm D4 brane to end on. 
Interestingly, 
the relative distance between two kinks corresponds to 
the size moduli of the single semi-local vortex. 
Moreover we can easily understand the small size limit 
of the configuration reduces to the ANO vortex with 
the ANO size $1/g\sqrt c$. 
Small lump singularity in the strong coupling limit 
$g \to \infty$ 
is resolved by the size of the ANO vortex for finite $g$, 
as discussed in Eq.~(\ref{eq:singular}). 
This example suggests the possibility to 
understand all moduli of vortices solely by kinky configurations, 
like Fig.~\ref{fig:tk=1}. 
We show that this is the case in the next section.

Before going that we make a comment. 
Taking T-duality further along the $x^3$-direction 
after compactifying that direction, 
the system is mapped to D1-D5 system 
where two NS5-branes are mapped to the asymptotically
locally Euclidean (ALE) geometry 
with ${\bf Z}_2$ singularity resolved by the string 
scale~\cite{Eto:2004vy}. 
(The four directions of D5 brane worldvolume 
perpendicular to D1 branes 
are divided by this ${\bf Z}_2$.)
In that brane configuration, 
D1 branes (instead of D2 branes in Fig.~\ref{fig:tk=1}) exhibit 
kinks interpolating separated D5 branes 
(instead of D4 branes in Fig.~\ref{fig:tk=1}).
In the discussion in the next section, 
we may consider either the D2-D4-NS5 system or the D1-D5-ALE system.

\section{Moduli Space of Vortices from Kinky D-branes}
\label{MSVW}
\setcounter{equation}{0}
In this section, we explore the relation between 
the moduli spaces of vortices and domain walls. 
In our previous paper \cite{Eto:2004vy} we realized 
domain walls in non-Abelian gauge theory by 
a kinky D-brane configuration. 
We would like to describe the moduli space of vortices 
on a cylinder without twisted boundary condition. 
However we put small mass splitting $m_A$ 
and analyze moduli in kinky D-branes 
after taking a T-duality along $S^1$. 
One advantage to do so is that 
the internal moduli of vortices, like orientational moduli 
of non-Abelian vortices, become visible. 
We also obtain the moduli space of vortices on a plane 
${\bf C}$, 
by taking the limit $\hat R \to 0$ of shrinking dual circle.

\subsection{Dimensionality}
\begin{figure}[h]
\begin{center}
\includegraphics[width=100mm]{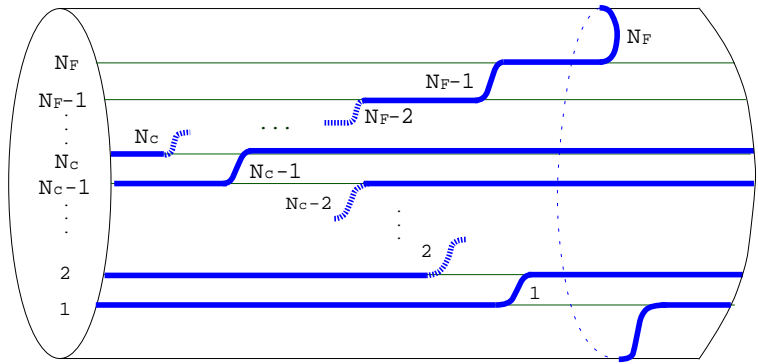}
\end{center}
\caption{ Brane configuration for one vortex.
\label{fig:onevortex}}
\end{figure}

First of all we discuss dimensionality of the moduli spaces. 
It is known that the dimension of 
the moduli space of the $k$ vortices is $2k\NF$ 
for arbitrary $\NC (\le \NF)$ \cite{Hanany:2003hp}. 
We now show that this can be seen from 
the T-dual picture of the vortex configurations. 
The T-dual picture of the one vortex configuration 
is given in Fig.~\ref{fig:onevortex}. 
This brane configuration can be decomposed into 
kinky D-branes exhibiting small kinks $\NF-1$ times and 
one large kink. 
They represent $\NF-1$ light walls and a heavy wall, 
respectively.  
Each wall have the two moduli parameters: 
one is for the position of walls and the other is for the phase. 
Therefore the dimension of the moduli space of one vortex is $2\NF$. 

\begin{figure}[h]
\begin{center}
\includegraphics[width=100mm]{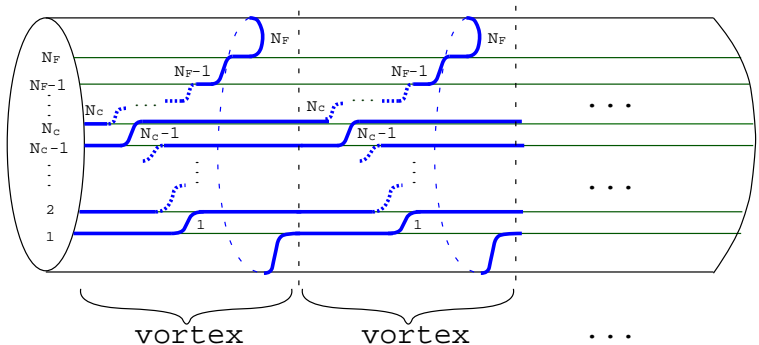}
\end{center}
\caption{ Brane configuration for $k$ vortex. 
\label{fig:kvortex}
}
\end{figure}

The T-dual picture of the $k$ vortices contains $k$ units 
of one vortex configuration as shown in Fig.~\ref{fig:kvortex}. 
In this brane configuration, 
there are $k\NF$ walls and thus we conclude that 
the dimension of the moduli space is $2k\NF$. 

We also find non-normalizable zero modes from these figures. 
In the limit of $m_A \to 0$ (without the twisted boundary 
condition), all D4 branes coincide and 
the moduli 
$G_{\NF,\NC} \simeq SU(\NF)/[SU(\NF-\NC) \times SU(\NC) \times U(1)]$
appear at both infinities of $x^1 \to \pm \infty$ from the freedom 
of D2-branes on D4-branes and become non-normalizable.

\subsection{Single non-Abelian vortex from kinky branes}
Next, we explore more concrete relations 
between the moduli spaces of vortices and walls. 
First, let us discuss a single vortex moduli space in 
the case of $\NF = \NC = 2$. 

The moduli matrix for this case on ${\bf C}$ is given 
by \cite{Eto:2005yh}
\beq
H_0= \bpm 
   1 & b \\ 
   0 & u-u_0 \epm   \hs{5} 
  b, u_0 \in \mathbf C \hs{5} u_0 \not = 0.
\eeq
By using the $V$-equivalence relation (\ref{V-equiv-vol}), 
it can also be written as
\beq
H_0 = \bpm 
    u-u_0 & 0 \\ 
 \tilde b & 1 \epm 
\hs{5} \tilde b = \frac{1}{b}.
\eeq
The moduli parameter $u_0$ corresponds to the position 
of the vortex. 
The parameters $b$ and $\tilde b$ are the inhomogeneous 
coordinates of $\mathbf {C}P^1$. 
We thus have ${\bf C} \times {\bf C}P^1$ as found earlier 
in \cite{Auzzi:2003fs}.
For $\NF=\NC=2$, the moduli space of one vortex on 
the cylinder $\mathbf R \times S^1$ is given by 
\beq
\mathcal M_{\NC=\NF=2}^{k=1} 
= {\mathbf C}^* \times \mathbf {C}P^1
\label{eq:vmoduli}
\eeq 
with ${\bf C}^* \equiv {\bf C} - \{0\} 
\simeq {\bf R} \times S^1$. 

\begin{figure}[h]
\begin{center}
\includegraphics[width=50mm]{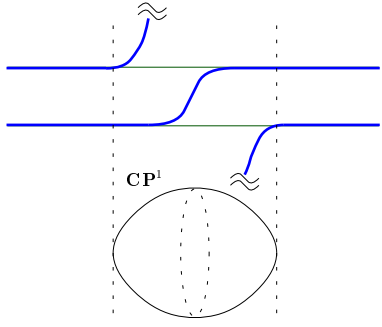}
\caption{Brane configuration for one vortex}
\label{fig:2N}
\end{center}
\end{figure}

\begin{figure}[h]
\begin{center}
\begin{tabular}{cc}
\includegraphics[width=50mm]{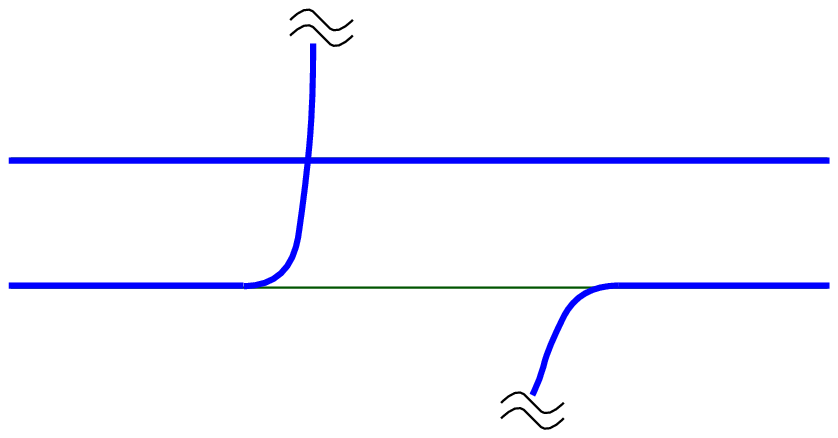} &
\includegraphics[width=50mm]{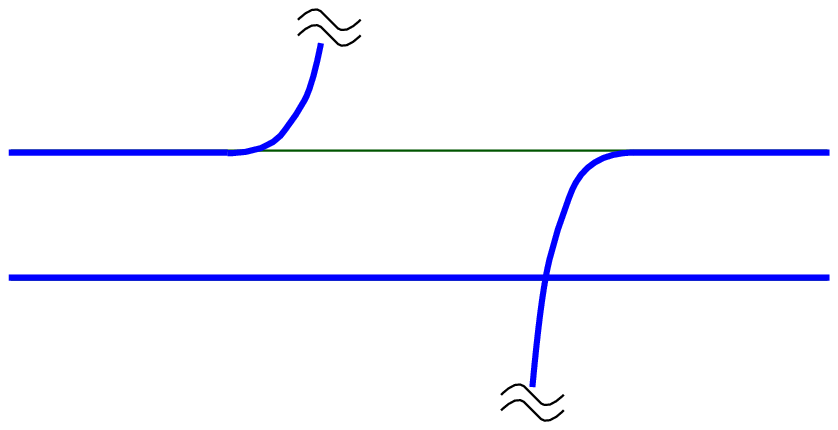} \\
(a) $b=0$ ($I=+1$) & (b) $\tilde b = 0$ ($I=-1$) 
\end{tabular}
\end{center}
\caption{The $b=0$ and $\tilde b=0$ limits.}
\label{fig:reconnection}
\end{figure}

Let us see this moduli space in terms of the T-dual picture. 
The brane configuration for the T-dual picture is given 
in Fig.~\ref{fig:2N}. 
There are two walls in this configuration 
and each wall has two moduli parameters. 
The position and the phase of the larger wall 
correspond to the position of the vortex 
in the $x^1$ direction and $x^2$ direction, respectively. 
The smaller wall also has two moduli parameters, 
that is the position and the phase. 
However, the position of this wall is confined inside 
the larger kink from the both sides, and 
hence the position modulus becomes a line segment. 
At the boundaries of that line segment, 
the D2 branes are reconnected~\cite{Eto:2004vy}. 
As a result we obtain 
a single D2 brane winding once 
and a straight D2 brane 
as shown in Fig.~\ref{fig:reconnection}. 
Therefore the phase 
degree of freedom
of the smaller wall disappears at the boundaries. 
We thus have $\mathbf {C}P^1$
from the moduli parameters of the smaller wall, 
reproducing the orientational moduli $\mathbf{C}P^1$ 
in the moduli space of the single vortex, given in 
Eq.~(\ref{eq:vmoduli}). 
Interestingly the winding D2 brane represents the ANO vortex. 

For later convenience we introduce the intersecting number $I$ of 
kinky D-brane configuration~\cite{Eto:2004vy}. 
When there exist no crossing D2 branes we define $I=0$. 
When reconnection occurs once it becomes $I = \pm 1$ with  
the sign depending on which brane goes across which brane.
If the straight-going upper (lower) D2-brane in the left side 
crosses over the lower (upper) D2-brane we define $I=-1$ ($I=+1$). 
The configurations in 
Fig.~\ref{fig:reconnection} (a) and (b) have $I=+1$ and $I=-1$, 
respectively.

\begin{figure}
\begin{center}
\includegraphics[width=50mm]{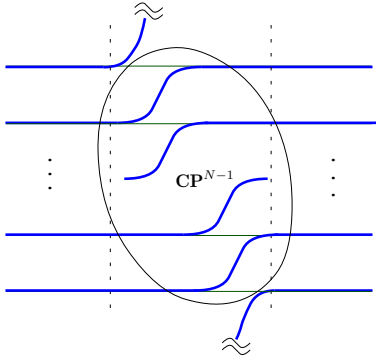}
\end{center}
\caption{Brane configuration for one vortex for $\NF=\NC=N$}
\label{fig:vortexN}
\end{figure}

The generalization to a single vortex moduli space for 
arbitrary $N \equiv \NC=\NF$ is straightforward. 
Fig.~\ref{fig:vortexN} shows the brane configuration 
for single vortex for $N$. 
The moduli space for internal walls is 
$\mathbf{C}P^{N-1}$ and 
this corresponds to the orientational moduli $\mathbf{C}P^{N-1}$ 
of one vortex, which comes from 
generators in $SU(N)/[SU(N-1)\times U(1)]$ broken 
by one vortex solution.
We again recover the single vortex moduli space 
${\bf C} \times {\bf C}P^{N-1}$ found in \cite{Auzzi:2003fs} 
in the limit of $\hat R \to 0$. 
Let us note that we cannot take the limit of 
$\hat R \rightarrow \infty$ to obtain meaningful wall 
configurations in the present case of 
$N_{\rm C}=N_{\rm F}$, since the larger wall becomes 
infinitely massive. 
We need to send the infinitely massive wall to infinity 
to decouple it, resulting in the unique vacuum without 
any walls, in conformity with our consideration in 
(\ref{eq:discrete_vacua}). @

\subsection{Double non-Abelian vortices from kinky branes}

We now discuss the multi-vortex moduli space 
by using the kinky D-branes. 
When $k$ vortices are well separated the moduli space 
tends to the symmetric product of $k$ single vortex moduli spaces, 
$({\bf C} \times {\bf C}P^1)^k/\Sym_k$, 
with $\Sym_k$ denoting the permutation group~\cite{Eto:2005yh}.
However the most important thing is how the moduli space looks  
when two or more vortices collide. 
This problem was discussed in \cite{Hashimoto:2005hi} 
in the case of $k=2$ for 
the moduli space proposed in \cite{Hanany:2003hp}, 
and has been clarified 
completely in \cite{Eto:2005yh} for general $N$ and $k$,  
by solving the BPS equations. 
Orbifold singularities of the symmetric product 
are resolved in the same manner with the Hilbert scheme 
for the case of instantons, resulting in a smooth moduli 
manifold.  
In this section, we discuss this problem by using 
the kinky D-brane configuration of the T-dual picture.

\begin{figure}[h]
\begin{center}
\begin{tabular}{cc}
\includegraphics[width=50mm]{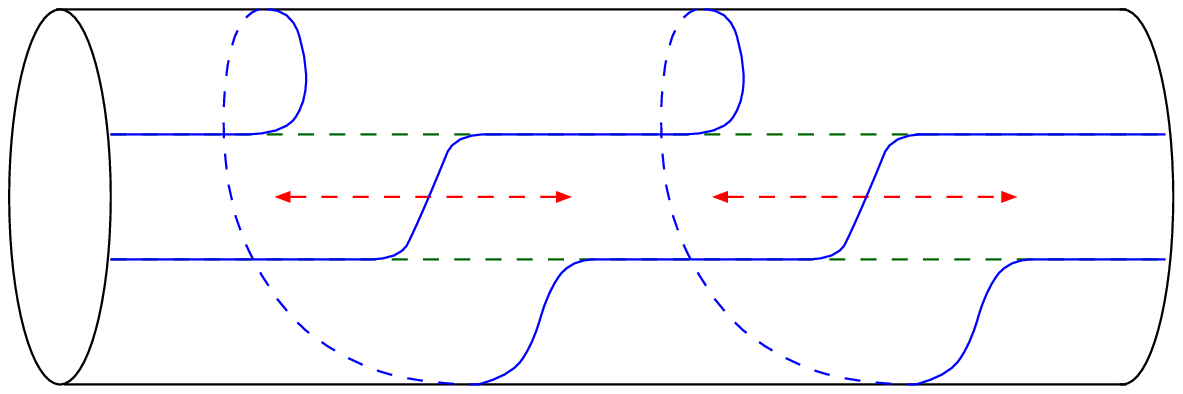} &
\includegraphics[width=50mm]{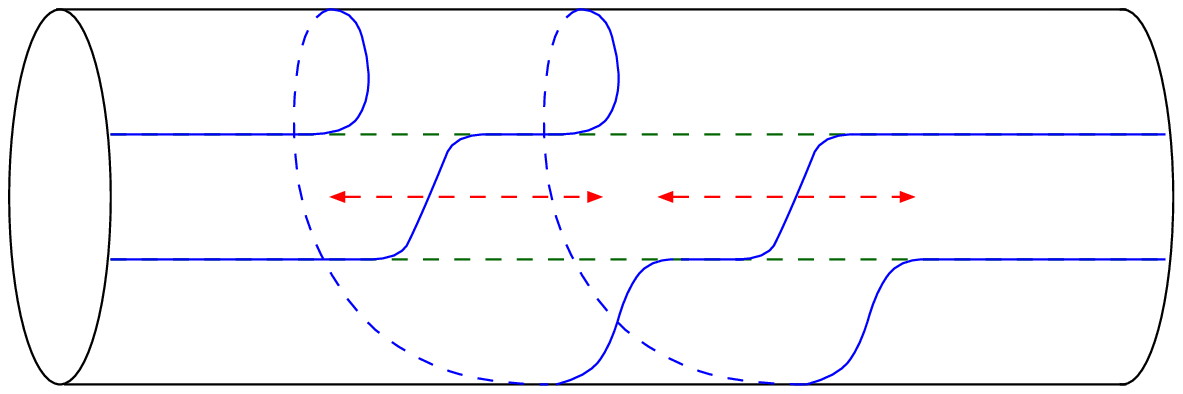} \\
(a) well separated vortices& (b) close vortices 
\end{tabular}
\end{center}
\caption{Brane configurations for two vortices.
The left-right arrows denote segments of movable region 
of small kinks.
\label{fig:2vortex}}
\end{figure}

For simplicity, we restrict ourselves to two vortices 
($k=2$) in the $N=2$ case, but we 
can easily generalize our consideration to arbitrary 
$k$ and $N$. 
 {
In this case the kinky D2-branes wind twice around $S^1$ in total. 
}
Depending on the relative position of two vortices, 
D-brane configurations are classified into two essentially 
different configurations as shown in Fig.~\ref{fig:2vortex}. 
In Fig.~\ref{fig:2vortex} (a) two vortices are well separated 
compared with the ANO vortex size $1/g\sqrt c$ 
and the configuration is just a sum of two configurations 
of one vortex in Fig.~\ref{fig:onevortex}. 
In  Fig.~\ref{fig:2vortex} (b)  the two vortices are 
close to each other as their distance is comparable with 
the vortex size $1/g\sqrt c$. 
The fact that the configuration of close vortices requires 
a separate treatment 
is the first sign that the moduli space 
is not just a symmetric product. 
 {
It is interesting to note that 
each D2 brane is configuration of 
a semi-local vortex 
in Fig.~\ref{fig:tk=1}~(b) or its inverse.
}

We can extract the moduli space of two vortices from 
these figures. 
When they are well separated 
there exist two copies of single vortices: 
the two large kinks give two ${\bf C}^*$ factors while  
the positions of the two small kinks are bounded from both 
sides (see Fig.~\ref{fig:2vortex}-(a))
just as in the single vortex case, 
giving two ${\bf C}P^1$ factors. 
Therefore as found in \cite{Hashimoto:2005hi,Eto:2005yh} 
asymptotic behavior of the moduli space of well 
separated vortices is found to be 
\beq
 \mathcal M_{\NC=\NF=2}^{k=2} \sim ({\bf C}^* \times {\bf C}P^1)^2 /\Sym_2.
  \label{eq:asympt}
\eeq 
However when the two vortices are close to each other,  
the situation becomes very different. 
In this case each small kink cannot move fully inside one 
large kink  unlike well separated vortices 
because its position is bounded by the identical D-brane 
on one side, where reconnection does not occur 
as seen in Fig.~\ref{fig:2vortex}-(b).  
Their phases still shrink there resulting in  {in} ${\bf C}P^1$'s 
again but their sizes are smaller than those in 
the case of the well separated vortices. 

\begin{figure}
\begin{center}
\includegraphics[width=120mm, angle = -90]{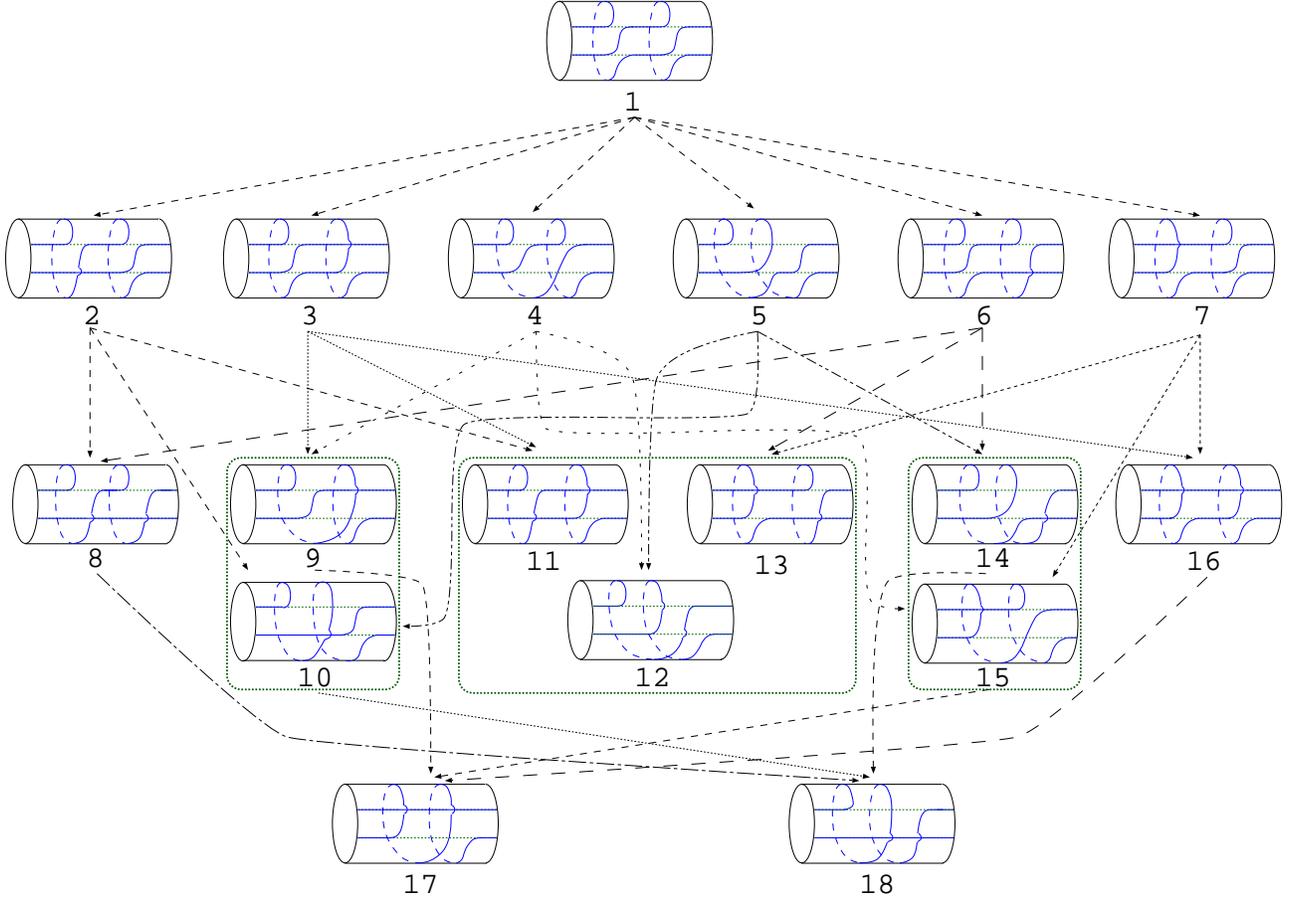}
\end{center}
\caption{ Brane configuration for two vortices. 
\label{fig:2vortex-all}}
\end{figure}

Let us discuss how two vortices collide in more detail.
All possible configurations for close vortices are drawn 
in Fig.~\ref{fig:2vortex-all} and are summarized in 
Table \ref{table:2vortex-all}.
\begin{table}
\begin{tabular}{c||c|cccccc|ccccccccc|cc}
 Config. & 
   1& 2& 3& 4& 5& 6& 7& 8& 9& 10& 11& 12& 13& 14& 15& 16& 17& 18 \cr \hline
 dim$_{\bf C}$ & 
 4 & 3 & 3 &3 &3 &3 &3 &2& 2& 2& 2& 2& 2& 2& 2& 2& 1&1 \cr
 $I$ &0& -1&-1&0&0&1&1& 2&-1&-1&0&0&0&1&1&-2&2&-2 \cr
\end{tabular}
 \caption{Moduli subspaces of two vortices.\label{table:2vortex-all}}
\end{table}
Each arrow in Fig.~\ref{fig:2vortex-all} 
denotes a suitable limit of one complex moduli parameter, 
resulting in a moduli subspace with complex dimension reduced 
by one. 
Hence the obtained subspace is a boundary of the space 
before taking the limit. 
Generic configuration of two vortices with full moduli 
dimension (complex dimension four)
is shown as the top figure. 
This configuration has six configurations 
(with the numbers from 2 to 7 in  Fig.~\ref{fig:2vortex-all}) 
on its boundaries, 
as listed in the second line. 
They are in complex dimension three and 
have the intersection number $I$ from $-1$ to $1$. 
All of them contain one ${\bf C}P^1$ factor and 
their moduli subspaces are like 
${\bf C}^* \times {\bf C}P^1 \times {\bf R}^+ \times S^1$.
Two vortices lie in the same $U(1)$ subgroup 
in the configurations 2 and 3 in  Fig.~\ref{fig:2vortex-all}.

In the third and fourth line 
the moduli boundaries reduced by one more complex dimension 
are listed (the configurations with the numbers from 8 to 16). 
They have the intersection number $I$ from $-2$ to $2$. 
Interestingly the two (three) configurations with 
$I= \pm 1$ ($I=0$) 
bounded by boxes in Fig.~\ref{fig:2vortex-all} 
are connected to each other without causing reconnection,
unlike the configurations in the second line. 
The three configurations with $I=0$ represent two vortices 
with different orientations. 
They can pass through each other without interaction, 
resulting in the moduli 
subspace $({\bf C}^*)^2$. 
More interesting phenomenon can be found in 
the $I=- 1$ case.
The movable position of the small kink in $I=-1$ 
is bounded from both sides like the single vortex case, 
resulting in a ${\bf C}P^1$ factor again. 
However the size of this ${\bf C}P^1$ is 
twice of that of the single vortex 
as was found in \cite{Hashimoto:2005hi}.\footnote{
See Eq.~(3.11) in Ref.~\cite{Hashimoto:2005hi} in which
such a moduli subspace was denoted by ${\cal M}|_{U(1), z=0}$.
}
In our configuration this phenomenon can be explained as follows. 
The large kinks denote two coincident ANO vortices. 
The size of two coincident ANO vortices 
is $\sqrt 2$ times that of single ANO vortex, 
if we assume that energy density is a step function 
with respect to a radius. 
This fact implies that the movable region of the small kink 
is $\sqrt 2$ times that of the single vortex 
($\frac{1}{g \sqrt c}$). 
This is effectively obtained by replacing $g^2$ by $g^2/2$.  
Since the K\"ahler potential for the single vortex moduli is 
given by $K \sim \frac{1}{g^2} \log (1+|b|^2)$, 
we have $K \sim \frac{2}{g^2} \log (1+|b|^2)$ for coincident two vortices. 
Thus the size of orientational modulus in the internal space has been 
explained by the size of the ANO vortices in the real space. 
The configurations with $I=+1$ also have a ${\bf C}P^1$ factor. 
In the configuration with $I = \pm 2$ the identical D2-brane 
winds twice around $S^1$ with the rest remains straight. 
These two vortices lie in the same $U(1)$ subgroup, 
and the configuration is the same with the case of two ANO vortices, 
resulting in the moduli space $({\bf C}^*)^2/\Sym_2$.

In the bottom line two configurations are drawn. 
Apparently both of them represent two coincident ANO 
vortices, resulting in the modulus ${\bf C}^*$. 

To summarize, the moduli subspaces with complex dimension 
$d$ have been obtained as
\beq
 && {\mathcal M}|^{d=4} 
    \simeq ({\bf C}^* \times {\bf C}P^1)^2 / \Sym_2,  \nonumber\\
 && {\mathcal M}|^{d=3}_{I=0,\pm 1} 
    \simeq {\bf C}^* \times {\bf C}P^1 \times {\bf R}^+ \times S^1, \nonumber\\
 && {\mathcal M}|^{d=2}_{I=0} 
    \simeq ({\bf C}^*)^2 , \quad
  {\mathcal M}|^{d=2}_{I=\pm 1} 
    \simeq {\bf C}^* \times {\bf C}P^1 , \quad
  {\mathcal M}|^{d=2}_{I=\pm 2} 
    \simeq ({\bf C}^*)^2/\Sym_2 , 
\nonumber \\ 
 && {\mathcal M}|^{d=1}_{I= \pm 2} 
    \simeq {\bf C}^* .
\eeq
 { 
The first one denotes an open set of the symmetric product 
removing orbifold singularity.
By connecting the rest of the moduli subspaces together 
and placing it to cover the hole of ${\mathcal M}|^{d=4}$
we obtain the two vortex moduli space ${\cal M}^{k=2}_{N=2}$. 
We conclude in this section 
that kinky configurations (in the large radius limit) 
give a method alternative to the one in \cite{Eto:2005yh} 
to obtain the smooth full moduli space,  
but not just the moduli subspace (\ref{eq:asympt}) for separated vortices. 
}

%

\section{Conclusion and Discussion}\label{Disc}
 {
We have established the duality between 
domain walls and vortices. The duality has been 
investigated from two points of view: field theory and
D-brane configurations. In terms of field theory, we 
have constructed vortices on cylinder $\mathbf R \times S^1$
as periodically arranged vortices on $\mathbf C$.
Then we have found that in the case of $\NC=1, \NF=2$, 
a pair of wall-like objects appears when the size
 of the vortex becomes larger than the period of 
the cylinder as shown in Fig.~\ref{fig:profile}. 
We have also obtained usual domain walls by introducing 
a twisted boundary condition and found that a vortex
can be decomposed into $\NF$ domain walls.
We have explained these phenomena as a T-duality between
the D-brane configurations of vortices (Fig.~\ref{fig:vortices})
 and domain walls (Fig.~\ref{fig:wall}). 
By using this duality, we have found some correspondence 
between the moduli space of non-Abelian vortices
and kinky D-brane configurations 
for domain walls. The dimensionality 
of the vortex moduli space can be calculated 
from its dual kinky brane configuration. 
The moduli spaces of single and double non-Abelian 
vortices have been explored in terms of 
the kinky brane configurations.
}

Here we give several discussions. 

\underline{Vortices on a torus}

In this paper, 
we have not considered two T-dualities 
along directions orthogonal to identical vortices. 
In order to do that, we have to consider vortices on a 
torus $T^2$ 
(see, e.g., Ref.~\cite{Gonzalez-Arroyo:2004xu} for 
analysis of the ANO vortices on $T^2$). 
This should be useful for construction of vortices 
if we remember that instantons on $T^4$ 
are related with the ADHM construction of instantons 
(the Nahm transformation). 
We expect that a reminiscent of the ADHM construction 
is obtained for vortices by taking T-dualities twice. 
This remains as an important future work. 

\medskip
\underline{T-duality between $1/4$ BPS composite states}

A T-duality between a $1/4$ BPS state of instantons 
as vortices in the worldvolume of a single vortex in $d=6$ 
($d=5$) gauge theory with massless hypermultiplets 
and $1/4$ BPS state of monopoles as kinks in it in 
$d=5$ ($d=4$) gauge theory 
with hypermultiplets with real masses   
was established previously in \cite{Eto:2004rz} 
through 1/4 BPS calorons in it.  
This analysis was done in the level of 
the effective theory, the ${\bf C}P^{N-1}$ model, 
on the vortex, 
which is completely parallel to the one in this paper. 

On the other hand, domain walls 
can make a junction or more generally a web (or network) 
as a $1/4$ BPS state
in $d=4$ or $d=3$ 
gauge theory with complex masses for 
hypermultiplets~\cite{Eto:2005cp,Eto:2005fm,Eto:2005mx}. 
This $d=4$ theory 
can be obtained by two Scherk-Schwarz dimensional 
reductions from $d=6$ theory with massless hypermultiplets. 
Accordingly the domain wall junction carries, 
at the junction point, the junction charge interpreted as 
the Hitchin charge, which is obtained 
by two reductions of the instanton charge.  
In fact, there exists in $d=6$ 
instantons accompanied 
with vortices in two directions, 
say orthogonal to $z=x^1 + i x^3$ and 
$w = x^2 + i x^4$ \cite{Eto:2004rz}. 
Taking a T-duality along $x^4$ 
this system becomes 
a $1/4$ BPS state of monopoles, vortices (on $z$) and 
domain walls (orthogonal to $x^2$) 
in $d=5$ theory with real masses for 
hypermultiplets~\cite{Isozumi:2004vg}.
Taking a T-duality once again along $x^3$ 
the system becomes 
$1/4$ BPS webs of domain walls 
in $d=4$ theory with complex masses for hypermultiplets. 
Thus all $1/4$ BPS states analyzed so far are 
connected by T-duality maps established in this paper. 

Another series of 1/4 BPS equations and the unique set of 
1/8 BPS equations 
have been found in \cite{Lee:2005sv,Eto:2005sw}. 
In particular the set of 1/8 BPS equations is obtained by 
considering vortices living on all possible two-dimensional 
planes orthogonal to 
each other in $d=6$ (in addition to instantons), 
and all 1/2, 1/4 and 1/8 BPS equations 
in dimensions less than $d=6$ are obtained by 
ordinary and/or the SS dimensional reductions \cite{Eto:2005sw}. 
Therefore they must be related through T-duality discussed 
in this paper. 
The brane construction of configurations for these newly 
found equations 
and T-duality among them remain as a future problem.

\medskip
\underline{Similarity with dyonic instantons as a supertube}

Dyonic instantons in $d=5$ gauge theory 
with sixteen supercharges 
are instantons with electric flux~\cite{Lambert:1999ua}. 
They can be interpreted as a supertube \cite{Mateos:2001qs} 
as follows \cite{Bak:2002ke}. 
The $d=5$, $U(2)$ gauge theory is realized on 
the world volume on two D$4$-branes. 
Dyonic instantons in this theory 
can be interpreted as 
a tubular D$2$-brane (a supertube) connected 
to the two separated D$4$-branes, 
carrying charges of D$0$-branes 
and fundamental strings. 
A ring made of  
ending points of D$2$-branes on a D$4$-brane 
is regarded as a loop of a monopole-string, 
with the total monopole charge being zero. 
It is known that this ring   
can be deformed to arbitrary shape 
by changing moduli \cite{Mateos:2002yf}. 
As an extreme case it can be deformed to a set of parallel 
monopole-sting and anti-monopole-string, where both of 
them are BPS preserving the same supercharges and 
therefore the total system is BPS.  
This situation is completely parallel to 
our analysis in this paper. 
Namely,  the shape of semi-local vortex 
can be deformed by changing moduli to 
a set of parallel domain wall and anti-domain-wall, 
both of which are BPS. 
 {
However we cannot deform the shape of the monopole-loop 
arbitrarily in our case. 
To overcome this, 
we should introduce time-dependence and 
replace lumps and walls by 
so-called Q-lumps \cite{Leese:1991hr} 
and Q-walls \cite{Abraham:1992vb}, respectively,  
which carry electric charges.  
Then we should be able to change the shape 
of the monopole-loop arbitrarily, 
and the situation becomes the same with 
the case of the supertube completely.\footnote{
 {
A similar discussion was made by Kimyeong Lee 
in his talk in the workshop at Kyoto in December 2005 \cite{Kim:2006}.
}
}
These new supertubes in the vortex/wall system 
are ``half" of the supertubes as dyonic instantons 
in the instanton/monopole system. 
We expect that this half property should be explained 
by putting NS5-branes in the brane configuration, 
like the relation 
between domain walls and monopoles \cite{Hanany:2005bq}. 
}


\underline{Similarity with tachyon condensation.}

In this paper, we see annihilation phenomenon of the wall 
and anti-wall pair, which leaves vortex after annihilation. 
The wall and anti-wall pair does not mean BPS and anti-BPS 
pair since the identical fraction of supersymmetry 
in the system is always preserved during 
the annihilation process. 
It is, however, reminiscent of the brane and anti-brane 
pair annihilation in string theory, where lower dimensional 
BPS branes survives after tachyon condensation. 
(See for review \cite{Harvey:2001yn}.)
The difference between the BPS and non-BPS system in 
superstring theory exists in the projections on the 
fermions like the relation between Type II and Type 0 
theory, but essential (bosonic) dynamics should be in 
common with each other.
Indeed, recently,
it is shown in \cite{Hashimoto:2005yy, Asakawa:2006}
that the ADHM/Nahm construction of instantons/monopoles has very
good interpretation in terms of the tachyon condensation
and the formulation can be extended to various dimensions and other
solitons.
Although we apply the vortex/wall system to only the 
BPS soliton system in this paper,  
we expect that our study sheds new lights also 
on the tachyon condensation 
of the non-BPS or non-commutative solitons.

\subsubsection*{Acknowledgements}

 {
We would like to thank Kimyeong Lee 
and other participants of the workshop ``Fundamental Problems 
and Applications of Quantum Field Theory" 
held at Yukawa Institute of Theoretical Physics 
in 19--23 December 2005. 
}
This work is supported in part by Grant-in-Aid for 
Scientific Research from the Ministry of Education, 
Culture, Sports, Science and Technology, Japan No.17540237 
(N.~S.) and 16028203 for the priority area ``origin of 
mass'' (N.~S.). 
The work of M.~N. and K.~Ohashi (M.~E. and Y.~I.) is 
supported by Japan Society for the Promotion 
of Science under the Post-doctoral (Pre-doctoral) Research 
Program. K.~Ohta is supported in part by 
Special Postdoctoral Researchers Program at RIKEN. 



\end{document}